\documentclass[12pt,a4paper,notitlepage,twoside]{article}
\usepackage{color,soul,amsmath,bbm,dsfont, mathrsfs,xcolor,setspace}

\usepackage[top=1in, bottom=1in, left=1in, right=1in]{geometry}

\normalfont

\usepackage{amsmath,amsthm,amsfonts,tikz,pgfplots,verbatim}
\usepackage{fancyhdr}
\usepackage{graphicx,caption}
\usepackage{subcaption}
\usepackage{relsize}
\usepackage{soul}

\usetikzlibrary{calc}

\usepackage{cite}

\tikzset{axis line style/.style={thin, gray, -stealth}}

\newcommand{\hlc}[2][yellow]{ {\sethlcolor{#1} \hl{#2}} }

\let\oldsqrt\sqrt
\def\sqrt{\mathpalette\DHLhksqrt}
\def\DHLhksqrt#1#2{%
\setbox0=\hbox{$#1\oldsqrt{#2\,}$}\dimen0=\ht0
\advance\dimen0-0.2\ht0
\setbox2=\hbox{\vrule height\ht0 depth -\dimen0}%
{\box0\lower0.4pt\box2}}

\title{{\Large Extrapolating the term structure of interest rates with parameter uncertainty}}
\author{ {\normalsize Balter, Anne}\footnote{Phone: +31433884962. Email: a.balter@maastrichtuniversity.nl}   \\ {\small Maastricht University}\footnote{Tongersestraat 53, 6211LM, Maastricht, The Netherlands} \and
{\normalsize Schotman, Peter}\footnote{Phone: +31433883862. Email: p.schotman@maastrichtuniversity.nl}  \\ {\small Maastricht University} \and
{\normalsize Pelsser, Antoon}\footnote{Phone: +31433883899. Email: a.pelsser@maastrichtuniversity.nl}   \\ {\small Maastricht University}}
\date{\small This version: December 18, 2013}

\begin{document}


\maketitle \thispagestyle{empty} 

\begin{abstract}
\noindent
Pricing extremely long-dated liabilities market consistently deals with the decline in liquidity of financial instruments on long maturities. The aim is to quantify the uncertainty of rates up to maturities of a century. We assume that the interest rates follow the affine mean-reverting Vasicek model. We model parameter uncertainty by Bayesian distributions over the parameters. The cross-sectional and time series parameters are obtained via the restricted bivariate VAR(1) model. The empirical example shows extremely low confidence in long term extrapolations due to the accumulated effect of the mean-reversion`s behaviour close to the unit root.
\end{abstract}


\newpage

\clearpage


\newpage
\section{Introduction}
Pricing extremely long-dated liabilities market consistently faces the difficulty of pricing in an incomplete market. The market in derivatives is incomplete since the liquidity of financial instruments declines over time. Long-dated liabilities have to be priced by pension funds and life insurance companies since the life expectancy goes beyond the maturity period of liquid assets. 
Therefore a method of extrapolating yield curves  far into the future is what we look for in this paper. Up to about two or three decades the Euro market is liquid, while rates with maturities up to a century are needed by pension funds. As funds are obliged to calculate the proper present value of the outstanding liabilities to determines the `health' of the fund. Currently the interest rates are low, what causes low funding ratios and bears a lot of concern to all generations. Therefore we investigate the methodology of extrapolating the term structure of interest rates on the very far end with a focus on the size of the uncertainty.

\vspace{5mm}
\noindent
Due to their tractability affine term structure models (ATSM) are widely used by both academics as well by practitioners in finance. Since we are interested in the long end of the curvature the level factor is the dominant one that determines the shape. Economic theory and historical data underscore a recurring pattern that high rates move downwards and low rates increase both to a constant level. The Vasicek model is based on these movements and is what we use for long term maturities. The parameter that pushes values towards a constant ultimate level has a large influence on the extrapolation of interest rates with long horizons. Therefore we are interested in the absolute value and the uncertainty of this parameter. Most literature looks at yields with maturities up to 10 years, a period in which the effect is not of high concern. However, rates in the range between 50 and 100 years, which are standard maturities needed by pension funds, are mostly determined by the mean-reversion parameter. Data shows that the mean-reversion is low, approaching the unit root. Also, for low values of the mean-reversion the ultimate interest rate shows to be extremely uncertain. Since both parameters are crucial for extrapolation we are interested in the total effect of the mean-reversion in extrapolations.

\vspace{5mm}
\noindent
We examine parameter uncertainty in the Vasicek model (1977)\nocite{vasicek1977equilibrium}. A Bayesian interpretation of the problem is applied since the uncertainty is applicable by defining a distribution over the parameters. Calculating the certainty equivalents of accumulated short rates following the Vasicek model after the last liquid point, is the methodology used for extrapolation of interest curves.
We determine the posteriors on a data set consisting of long term bond rates and a set of constructive priors.
To stress the extremely long end of the curvature also the liquid input maturities are chosen to have medium to long horizons compared to standard term structure modelling.

\vspace{5mm}
\noindent
We compare the Vasicek extrapolation with the Nelson-Siegel (NS) method and the Smith-Wilson (SW) method in combination with an Ultimate Forward Rate (UFR). In terms of volatility we can rank the NS model cross sectionally as extremely volatile, whereas the SW is has no uncertainty at the long end by construction and the Vasicek model is between. 

\vspace{5mm}
\noindent
The setup of this paper is as follows. Section 2 describes the theoretical Affine model. Section 3 describes the data. Then we further specify the econometric model and decompositions of the covariance matrix in Section 4. In Section 5 the frequentistic conditional Maximum Likelihood method is applied and in Section 6 the Bayesian approach is explained. Whereas an empirical application of the Bayesian technique and a run of the model including a discussion of the results can be found in Section 7. Section 8 contains the extrapolation of the empirical application compared with the Nelson-Siegel and Smith-Wilson UFR method. Then a discussion and some robustness notes are made and lastly, the conclusion makes up Section 10.

\section{Affine Term Structure Model }
The Vasicek model describes a process that is autoregressive and converges to a long-term mean. It obtained much popularity since its practical economical application and its analytical tractability. Many valuations of asset pricing can be solved analytically under this model, i.e. pricing discount bonds, options on discount bonds, caps, floors, swaption. However, here we will fit the term-structure of interest rates under the assumption of parameter uncertainty which makes the problem only numerically solvable. For time steps $h$ the transition density of the Vasicek model is identical to the discrete time autoregressive process of order 1 (AR(1)),  hence the two processes will be used interchangeably both for mathematical and implementational convenience.

\vspace{5mm}
\noindent
Whether we work with the process of the short rates or with the process of the zero-rates is identical since a Vasicek for the one results in a one-to-one correspondance of a Vasicek model for the other due to affine relation. The mean reverting continuous-time stochastic differential equation (SDE) of the Vasicek model under the real-world probability measure $\mathbb{P}$ for the short rate $r$ is
\begin{equation}
dr_t = -\kappa(r_t-\mu)dt+ \sigma dW_t \nonumber
\end{equation}
We work with the following model, since data is observed in terms of zero-rates ($z_t$) 
\begin{equation}
\label{OU}
dz_t=-\kappa (z_t-m) dt+ \sigma b(\tau) dW_t
\end{equation}
By Eulers decomposition and the continuity corrections, the direct expression for the zero rate is
\begin{equation}
\label{AR1}
z_{t+h}=z_t - \frac{1-e^{-\kappa h}}{\kappa} h (z_t-m) +\sigma \sqrt{\frac{1-e^{-2\kappa h}}{2\kappa} }e_{t+h}
\end{equation}
where $e_t \sim N(0,1)$ is iid. 


\noindent
To forecast we switch to the risk-neutral probability measure $\mathbb{Q}$. Since the expected accumulation of short rates calculated as a discount factor corresponds to a zero-rate, an analytical formula expresses zero-rates with a maturity $s$ in terms of a shorter maturity $\tau$. Let the stochastic discount factor (SDF), the Radon-Nikodym derivative (RN), deflator or also known as the Pricing Kernel (PK) be defined by
\begin{equation}
\frac{d \Lambda}{\Lambda} = -r_t dt - \lambda_t dW 
\end{equation}
where
\begin{equation}
\lambda_t = \Lambda_0 + \Lambda_1 r_t
\end{equation}
which is used in essentially affine models among others by Duffee (2002)\nocite{duffee2002term}. This determines the affine relation of the natural logarithm of the bond price, rewritten in terms of yields as
\begin{equation}
z(\tau)=-\tfrac{1}{\tau}A(\tau)- \tfrac{1}{\tau}B(\tau)r_t
\end{equation}
The Fundamental Pricing Equations implies that both measures are related by (see Appendix \ref{App:AppendixA})
\begin{eqnarray}
\tilde{\kappa}= \kappa + \sigma \Lambda_1 \nonumber\\
\tilde{\mu} \tilde{\kappa} = \mu\kappa - \sigma \Lambda_0
\end{eqnarray}
where the tilde represents the risk-free measure $\mathbb{Q}$, and the variables without tilde come from the historical measure $\mathbb{P}$.
The process of the short rate can be expressed under both measures.


\vspace{5mm}
\noindent
Under the risk-neutral measure the transition from the short to the zero rates can be made by solving the expectation
\begin{equation}
\mathbb{E}_{\mathbb{Q}}\left[e^{-\int_{0}^{\tau} r_s ds}\right] = e^{-\tau z(\tau)}
\end{equation}
into
\begin{equation}
\label{y_t}
z(\tau) = {b}(\tau) \bigg[r_t - {\theta} \bigg] + {\theta} + \frac{1}{2} \tau \omega^2 {{b}(\tau)}^2 
\end{equation}
where \begin{eqnarray}
{b}(\tau) &=& \frac{1-e^{-\tilde{\kappa} \tau}}{\tilde{\kappa} \tau} \nonumber\\
{\theta} &=& \tilde{\mu} -\frac{\sigma^2}{2\tilde{\kappa} ^2} \nonumber\\
\omega^2 &=& \frac{\sigma^2}{2\tilde{\kappa} } \nonumber 
\end{eqnarray}
The function ${b}(\tau)$ quantifies the sensitivity of long-term yields with respect to the short rate $r$, $\omega^2$ is the unconditional variance of the short rate and for $\tau \rightarrow \infty$ the yield converges to $\theta$, the long-term mean which equals the risk-neutral mean of the short rate minus the infinite horizon convexity adjustment. All zero rates are a weighted average of the current short rate and the long term yield plus a convexity adjustment. The derivation is shown in Appendix B.

\vspace{5mm}
\noindent
The above formula can be used to express the dependence of two yields with different maturities. For $s>\tau$

\begin{equation}
\label{y_t2}
z(s) = \frac{{b}(s)}{{b}(\tau)} \bigg[z(\tau) - {\theta} \bigg] + {\theta} + \frac{1}{2} \omega^2 {b}(s) \bigg(s {b}(s) - \tau {b}(\tau)\bigg)
\end{equation}
We shall refer to this expression as the extrapolation method. The convergence speed $\frac{{b}(s)}{{b}(\tau)}$ represents the mean-reversion from some future yield compared to a quoted and liquid yield to move towards the long-term mean. It can also be interpreted as the relative volatility, $\frac{\textrm{vol}[z(s)]}{\textrm{vol}[z(\tau)]}=\frac{{b}(s)}{{b}(\tau)}$, this is a declining function from 1 to 0 for $s$ increasing.

\vspace{5mm}
\noindent
The aim is to determine long-term interest rates while accepting parameter estimation error. In classical econometrics the asymptotic distribution approximates a finite sample. Incorporating uncertainty by defining a distribution over the true parameters is a way to include uncertainty. Before we describe the model that adds the parameter uncertainty we first show the data briefly. After which we specify the model in more details and apply two different models that quantify the uncertainty. 

\section{Data}
\begin{figure}[h!]
\centering
\caption*{Euro swap rates}
\begin{subfigure}[b]{0.45\textwidth}
	\centering
	\includegraphics[width=\textwidth]{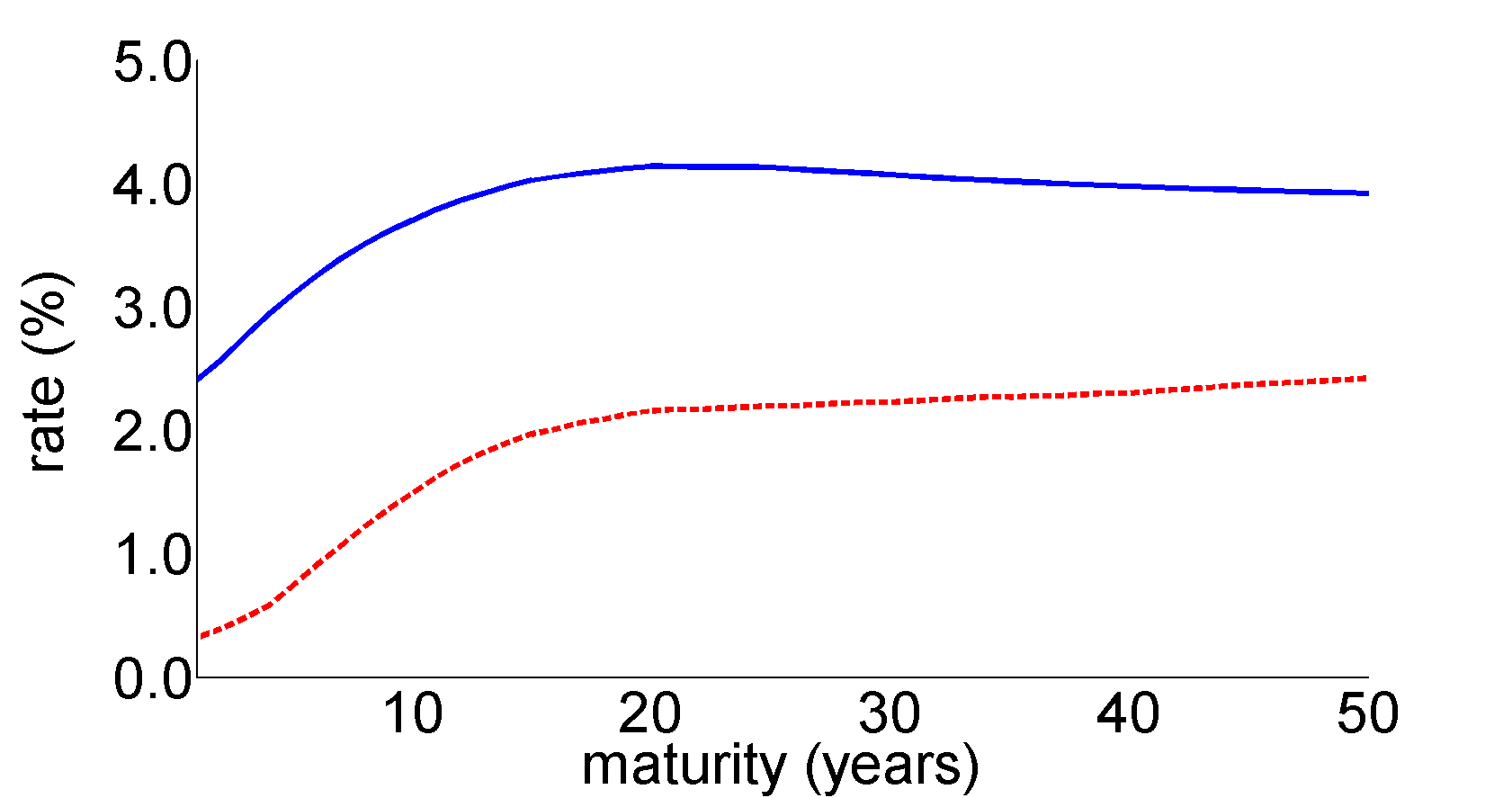}
	\caption{Levels}
	\label{fig:level}
\end{subfigure}
~
\begin{subfigure}[b]{0.45\textwidth}
	\centering
	\includegraphics[width=\textwidth]{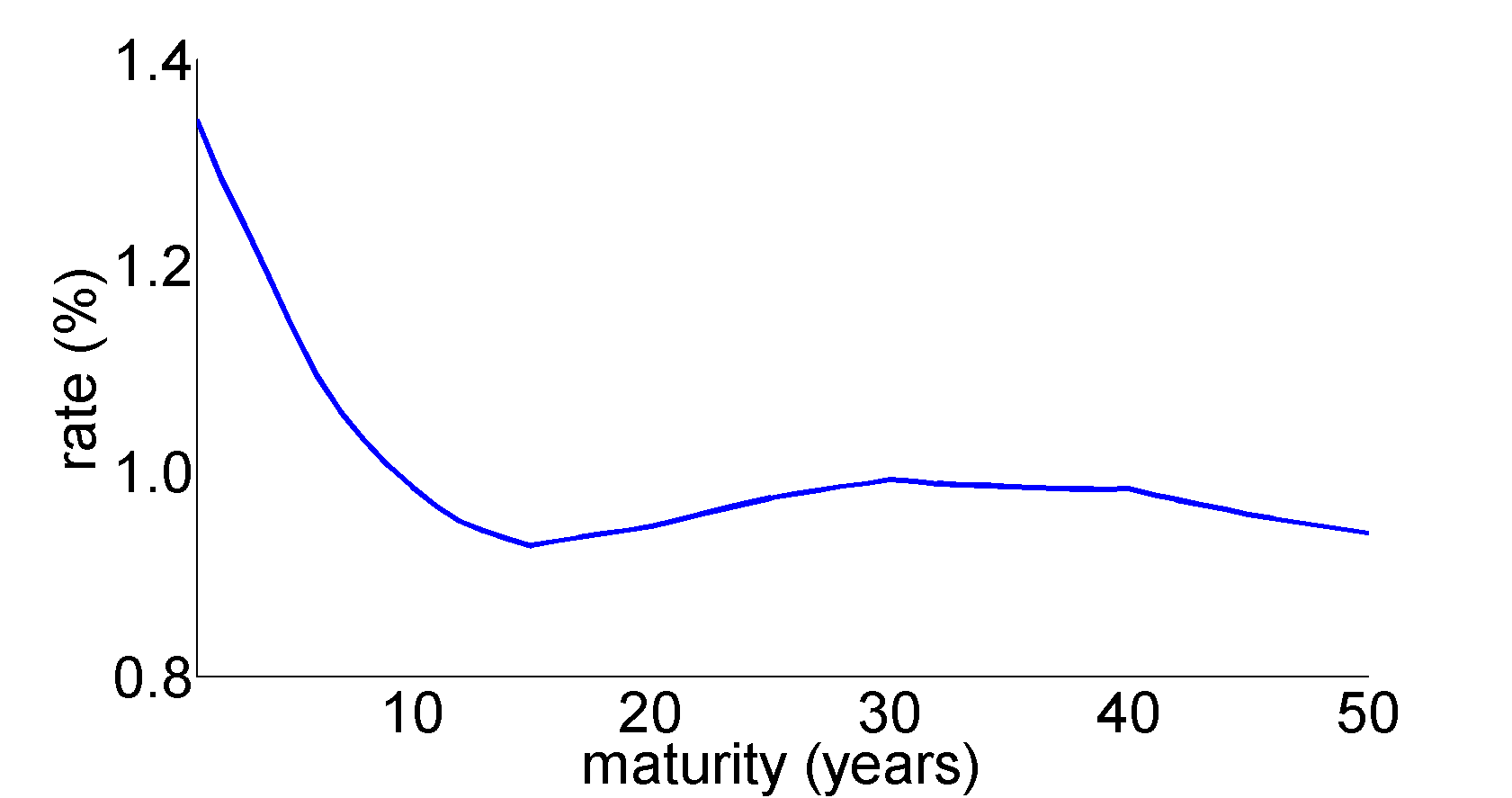}
	\caption{Volatility}
	\label{fig:vol}
\end{subfigure}
\caption{{\small \textit{The figure on the left shows the average term structure of interest rates with maturities of $1,2,3,...,50$ years and historically averaged over the period from January 2002 untill September 2013. The blue line shows the historic average whereas the dashed red line the term structure of Seotember 2013 shows. The figure on the right shows the average volatility for the same set of maturities and time series. The rates are given in percentages and the maturities in years.}}}\label{fig:Euro_swap_rate_September}
\end{figure}

Monthly zero-coupon Euro swap rates with maturities ranging from 1 to 50 years are used from the website of Bundesbank\footnote{From http:\//www.bundesbank.de/Navigation/EN/Statistics/Time\_series\_databases}. The sample period is from January 2002 to September 2013 resulting in 140 data-points per maturity. The average term structure has increasing yields until the 20-year maturity, after which it becomes slightly downward sloping for longer maturities. The initial hump shape for shorter to intermediate maturities can only be explained by a multiple factor model. However we are interested in long-dated maturities where the curve is smooth without humps. Henceforth a one-factor model like an AR process can capture this. Figure \ref{fig:Euro_swap_rate_September} shows the average term structure and the term structure from September 2013 which depicts the current situation of extreme low rates.

\noindent

\noindent
Figure \ref{fig:vol} shows an upward sloping pattern from a maturity of 15 years onwards, which is neither common in historical data nor caught by theoretical term structure models. The AR model, amongst all mean-reverting models, implies that the volatility curve is downward sloping for longer maturities. An explanation for this unexpected direction can be that very long-dated swap prices contain more noise because the market at this far end of the time line is illiquid. 

\begin{figure}[h!]
\centering
\caption*{Time series data}
\begin{subfigure}[b]{0.45\textwidth}
	\centering
	\includegraphics[width=\textwidth]{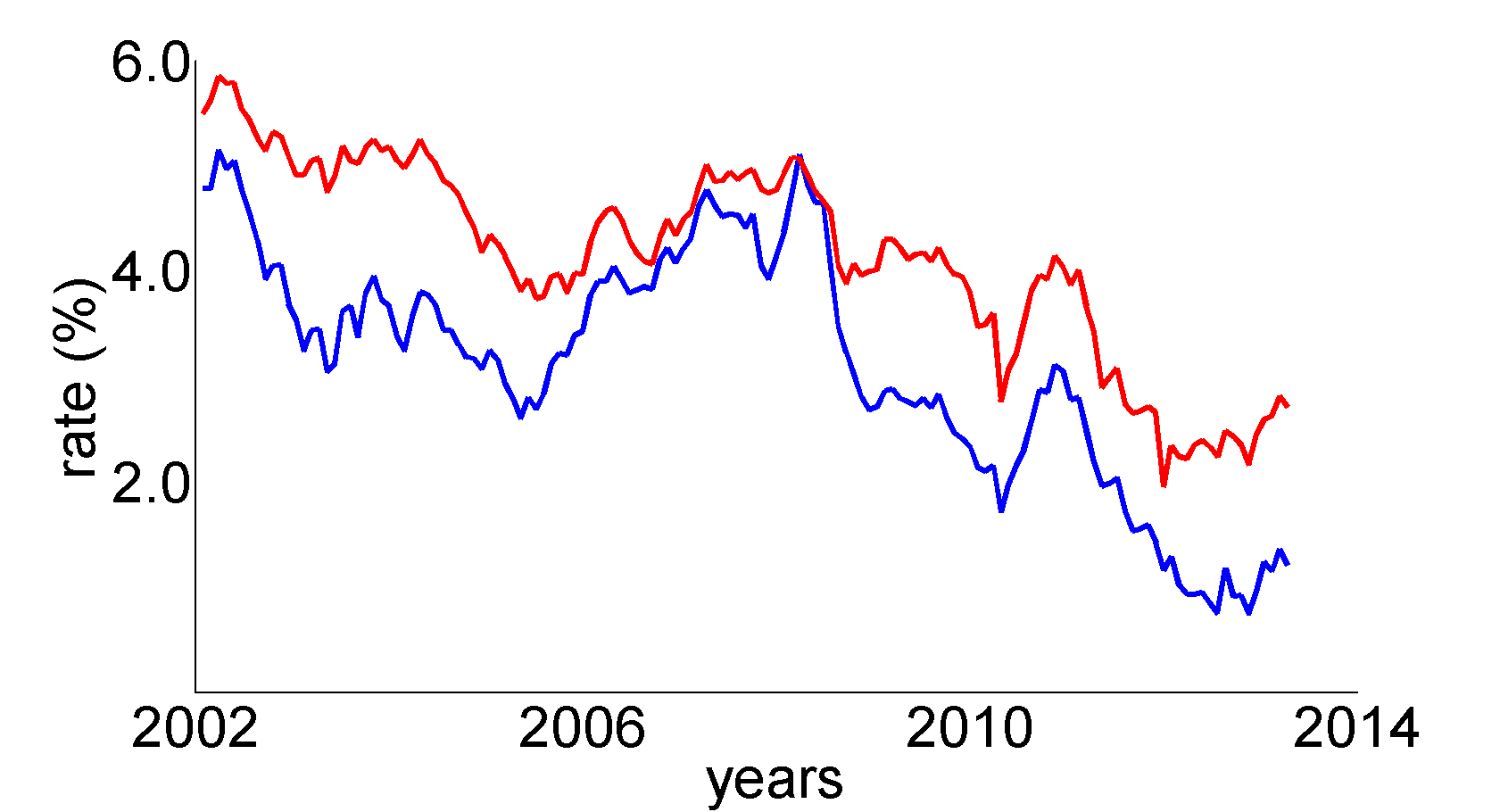}
\end{subfigure}
\caption{{\small \textit{The blue line is the time series path of the 5-year maturity swap interest rates from January 2002 untill September 2013 and the red line shows the historical development of the 20-year rates.}}}
\label{figure:historic}
\end{figure}

\vspace{5mm}
\noindent
The complete data set can be interpreted as panel data. Where we have time-series by considering a fixed maturity, resulting in a set of 140 historical observations of that maturity rate. And if a time-point in history is fixed, then a complete cross-sectional term structure from that period is found. 
Since we are intersted in the very long end of the curve we also use only relatively long maturities as input for the model. See the historical development of the 5 and 20 year interest rates in the period $[\textrm{Jan}, 2002: \textrm{September}, 2013]$ in Figure \ref{figure:historic}. 

\noindent

\section{Econometric model}
\label{section_var_cov_decomp}
\noindent
In order to calculate the long-term mean we need to know the process of the stochastic discount factor $\frac{d\Lambda}{\Lambda}$. A method to derive the long-term mean is to calculate the long-term mean and mean-reversion parameters under the risk-neutral and the physical probability measures. Since the relation between the two measures depends on the market price of risk, knowing $\lambda_t$ or $\tilde{\kappa}, \tilde{\mu}$ is equivalent. The latter method is what we will apply here, the restriction on the SDF is Cochrane and Piazzesi's (2009) approach\nocite{cochrane2009decomposing}. Joslin, Singleton and Zhu (2011)\nocite{joslin2011new} show that without restrictions on the risk pricing the historical based estimates do not add information onto the risk-neutral estimates. To derive the parameters under the cross-sectional measure $\mathbb{Q}$ historical data from at least two different maturities are needed. With a single interest rate time series it is impossible to identify the cross-sectional parameters. With multiple maturities the parameters are overidentified. The use of the discrete AR(1) process as the equivalent of the continuous Vasicek model can be extended to higher dimension, as such here the bivariate Vasicek model has a bivariate AR(1) analog. In this bivariate process the means of the two maturities can differ, while the mean-reversion parameter should be unique for the process and a unique one-dimensional variation is imposed.

\vspace{10mm}
\noindent
Henceforth, consider single factor model as following the VAR(1) process
\begin{equation}
\label{y_t_matrix}
\begin{bmatrix}
z_{t}(\tau_1) \\ 
z_{t} (\tau_2)
\end{bmatrix} =
\begin{bmatrix}
z_{t-h} (\tau_1)  \\ 
z_{t-h} (\tau_2)
\end{bmatrix} - a h 
\begin{bmatrix}
z_{t-h}(\tau_1 ) - m(\tau_1) \\ 
z_{t-h}(\tau_2 ) - m(\tau_2)
\end{bmatrix} + \sqrt{h}
\boldsymbol{\sigma}
\begin{bmatrix}
e_t^{(1)} \\
e_t^{(2)}
\end{bmatrix}
\end{equation}
\begin{equation}
\boldsymbol{Z}_t=\boldsymbol{Z}_{t-h}- a h(\boldsymbol{Z}_{t-h} - \boldsymbol{m})+\sqrt{h} \boldsymbol{\sigma}  e_t
\end{equation}
where $e_t^{(1)}$ and $e_t^{(2)}$ are from a bivariate standard Normal distribution with covariance matrix $\boldsymbol{\varSigma}_{\textrm{\tiny SIMUL}}=\boldsymbol{\sigma\sigma}'=\begin{bmatrix}
\sigma_{(11)} & \sigma_{(21)} \\ 
\sigma_{(21)} & \sigma_{(22)}
\end{bmatrix}$.
The continuity error corrections on $\kappa$ and $\sigma$ are  $a_{\textrm{DRAW}} = \frac{1-e^{\kappa h}}{\kappa} $ and $\sqrt{h}\boldsymbol{\sigma_{\textrm{DRAW}}}=\sqrt{\frac{1-e^{-2\kappa h}}{2\kappa}} \boldsymbol{\sigma}$.

\vspace{5mm}
\noindent
The theoretical model will have a covariance matrix implied by relation \eqref{y_t2},
\begin{equation}
\boldsymbol{\varSigma}_{\textrm{\tiny MODEL}} = \sigma^2
\begin{bmatrix}
{b}(\tau_1)^2 &  {b}(\tau_1) {b}(\tau_2) \\
 {b}(\tau_1){b}(\tau_2) & {b}(\tau_2)^2
\end{bmatrix}
\end{equation}
This matrix has $\textrm{rank}$ one. However there is some noise since yields are observed with error. The 5 parameters of the short rate Vasicek model, $\mu, \tilde{\mu}, \kappa, \tilde{\kappa}$ and $\sigma$ are overidentified. Real data will not exactly identify the theoretical matrix. Include this noise as either a correlation or as an extra noise term. A correlation coefficient will go to $\rho \rightarrow 1$ and a noise term will go to $\eta \rightarrow 0$ if the data behaves more and more like the one-factor model at hand. In empirical research both methods are adopted. De Jong (2000)\nocite{jong2000time} specifies a measurement error in his state-space model with multi-factors. Note that more maturities are needed to identify the parameters which are estimated by the use of the Kalman filter. In De Jong's paper (2000)\nocite{jong2000time} the one-factor model shows substantial misspecification of a general term structure. Including three factors (level, steepness and curvature respectively (Litterman and Scheinkman (1991))\nocite{litterman1991common}) seems to capture the movements of historical data best (Dai and Singleton (2000)\nocite{dai2000specification}). More specifically Litterman and Scheinkman (1991) showed that about 90\% of the variation can be explained by the first factor, however we are only interested in the extreme long-end of the term structure for which a single factor fits the needed characteristics. To stress the different fields of modelling among De Jong and Litterman and Scheinkman and this paper, both De Jong and Litterman and Scheinkman include short maturities where our model does not include rates below a maturity of 5 years.

\vspace{5mm}
\noindent
As the correlation decomposition and the noise decomposition turn out to be very similar, we shall decompose the covariance matrix in terms of an error component $\eta$. 
\begin{equation}
\boldsymbol{\varSigma}_{\eta}= \boldsymbol{\varSigma}_{\textrm{\tiny MODEL}} + \boldsymbol{I}\eta = \sigma^2
\begin{bmatrix}
{b}(\tau_1)^2 + \eta &  {b}(\tau_1) {b}(\tau_2) \\
 {b}(\tau_1){b}(\tau_2) & {b}(\tau_2)^2 + \eta
\end{bmatrix}
\end{equation}

\noindent
First $\tilde{\kappa}$ can be obtained numerically and based on this the other parameters are analytically solvable. 
\noindent
The nonnegativeness of $\tilde{\kappa}$ is ensured by
\begin{equation}
\label{restriction_kappa}
\tilde{\kappa} \geq 0 \Leftrightarrow \frac{\sigma_{(11)}-\sigma_{(22)}}{\sigma_{(21)}} \geq 0
\end{equation}
Since the condition that the numerator is larger than zero is imposed by the $\rho$-decomposition, $\sigma_{(11)} \geq \sigma_{(22)}$, we add the condition $ \sigma_{(21)} > 0$.

\noindent
Denote the expectation of the zero rates by $m(\tau_i)$. If we know the formulas $b(\tau_i)$ and the simulated values $m(\tau_i)$ then the bivariate process yields two equations for two unknowns, $\theta$ and $\mu$.  
\begin{equation}
\begin{bmatrix}
m(\tau_1) \\ 
m(\tau_2)
\end{bmatrix} =
\begin{bmatrix}
{b}(\tau_1) & 1-{b}(\tau_1) \\ 
{b}(\tau_2) & 1-{b}(\tau_2)
\end{bmatrix} 
\begin{bmatrix}
\mu \\
{\theta}
\end{bmatrix} + \frac{1}{2}{\omega}^2
\begin{bmatrix}
\tau_1 {b}(\tau_1)^2 \\
\tau_2 {b}(\tau_2)^2
\end{bmatrix}
\end{equation}

\noindent
As we already derived the relation between ${\theta}$ and $\tilde{\mu}$, also the implied 
\noindent
stochastic discount factor is known. Hence we end up with formulas for $\mu, \tilde{\mu}, \kappa, \tilde{\kappa},\sigma^2,$
\noindent
$\rho,\eta,\theta,\Lambda_0$ and $\Lambda_1$.

\section{Maximum Likelihood estimates}
\label{section_cMLE}
We can apply the conditional Maximum Likelihood Estimation (cMLE) to equation \eqref{y_t_matrix}. Note that we conditioned on the first observation. Under a classical interpretation the parameters are asymptotically Normal distributed with a mean equal to the cMLE and the variance obtained by the inverse of the negative expectation of the second order derivative. Simply maximizing the conditional log-likelihood function (see Appendix E) and retaining the Hessian matrix results in the asymptotic distributions of the parameters. The asymptotic variances of the decompositions is approximated by the Delta method.

\vspace{5mm}
\noindent
Applied to the Euro swap rates this results in the estimates observable in Table \ref{table:cMLE}. The point estimates shown in the first column are analytically obtained, where $\tilde{\kappa}$ is stated by an implicit function. Due to the nonlinearity the variances are obtained by the Delta method, shown in the second column.

\begin{table}[!h] 
\caption{Parameters noise decomposition based on cMLE} 
\begin{center} 
  \begin{tabular}{ l  l  l} 
    \hline 
& Estimate &Standard Error\\ 
\hline 
$\kappa$ &   0.2056  &   0.1083 \\ 
 $\tilde{\kappa}$ &   0.0201 &     1.4411\\ 
 $\mu$ &   0.0103   &   1.0855\\ 
 $\tilde{\mu}$ &   0.1338   &  37.4293\\ 
 $\theta$ &    0.07545   &   27.06 \\ 
 $\Lambda_0$ &   -0.0817  &   57.1148\\ 
 $\Lambda_1$ & -27.0356 &  210.5676 \\ 
 $\sigma^2$ &    4.710$\times 10^{-5}$  &     1.864$\times 10^{-3}$ \\ 
 $\eta$ &    1.086$\times 10^{-5}$ &    2.138$\times 10^{-3}$ \\ 
    \hline 
\end{tabular} 
\end{center} 
\caption*{\small {\textit{Conditional Maximum Likelihood applied to bootstrapped zero rates from the Euro swap rates with maturity 5 and 20 years. The standard errors are obtained via the Delta approximation.}}}
\label{table:cMLE}
\end{table} 
\noindent 


\vspace{5mm}
\noindent
The uncertainty of the mean-reversion is enormous. No meaningful conclusions can be extracted from this table as all important parameters include high variation. The common belief of positive average mean-reversion and means are not rejected by the cMLE, but neither negative ranges. The impact of the uncertainty is not extended in the literature so far since the influence is not that dramatic yet if one is interested in forecasts on a limited horizon. However, if maturities of extrapolation are in the range of 50 years and more, even up to 100 year, the effect is large. If the mean-reversion ($\tilde{\kappa}$) goes to zero, the ultimate forward rate ($\theta$) goes to minus infinity which lacks a possible economic explanation. Therefore we like to quantify the uncertainty of the mean-reversion parameter as the sensitivity of the zero-rate is hereby determined.

\vspace{5mm}
\noindent
As a general discussion, the uncertainty of all parameters is rather large and all intervals contain negative values. A comparison with the Bayesian method follows in Section 6. The large standard deviations result in unrealistic intervals for extrapolations on long horizons.

\section{Bayesian approach}
By considering parameter uncertainty as a point of research a Bayesian viewpoint fits to this problem since Bayesians specify probability densities over parameters. And restrictions are easily implemented in the algorithm. The Gibbs sampler is used since the likelihood function of the data is Normal and accordingly the separate posterior distributions are identifiable. For a Bayesian background, choices of priors and ways of generating posteriors see Bauwens, Lubrano and Richard (1999)\nocite{bauwens1999bayesian}. 
By Gibbs Sampling we can sample and find the posterior densities numerically using Markov Chain Monte Carlo simulations (MCMC). Together these draws will converge to the joint distribution. For one parameter the posterior distribution is known conditional on the other parameters. Iteratively one draw will be made conditional on all other parameters, next the other parameter is drawn conditional on all current values for the other parameters, et cetera. By Bayes rule the conditional posterior distributions can be derived. 

\vspace{5mm}
\noindent
The mean reversion parameters is assumed to be positive in accordance with economical belief. When interest rates are high the mean-reversion parameter pulls the rates down in correspondence with economical behavior since in times of high rates the economy tends to slow down which decreases investments which decreases demand for money and this triggers a decline of the interest rates. On the other hand if interest rates are low, investing is relatively cheap which causes an increase of interest rates due to a higher demand of money. The mean-reversion parameter accounts for these movements and makes this a useful and realistic model. Therefore the prior of $a$ is the truncated Normal distribution
\begin{equation}
f(a) \sim TN( \mu_a , \tau_a^2)
\end{equation}
with $\mu_a=0, \tau_{a}=0.2$. Note that the prior mean and standard deviation are $\mathbb{E}[a]=0.16$ and $\sqrt{\textrm{var}[a]}=0.12$ by this choice.

\vspace{5mm}
\noindent
Furthermore we also assume the long term mean of the zero rates to be positive. Moreover we do not put any dependence between the two sets of maturities upfront. The prior of $\boldsymbol{m}$ is a two dimensional truncated Normal distribution
\begin{equation}
f(\boldsymbol{m} ) \sim TN_2\left( \boldsymbol{\mu_m}, \boldsymbol{\Omega_m} \right)
\end{equation}
with $\boldsymbol{\mu_m}=[ -0.923, -0.923]'$ and $\omega_{m(1,1)} = \omega_{m(2,2)} = 0.2$ and $\omega_{m(1,2)}=\omega_{m(2,1)}=0$ implying the mean to be  $\mathbb{E}[m]=0.04$ and $\sqrt{\textrm{var}[m]}=0.039$. The difference between the hyperparameters and the mean and variance are due to the truncated part of the distribution, the negative part of the standard Normal distribution is left out in the mean. The range of both priors include a realistic, above zero, and large set of different priors. The implementation of a truncated Normal is simply generated by drawing from a Normal where one rejects the negative draws. The rate of acceptance will be extremely low if the mean reversion parameter is close to the unit root. This results in drawings for $\boldsymbol{m}$ from close to the prior distribution with a negative hyperparametric mean. If the standardized truncation parameter is above a certain treshhold exponential rejection sampling (Geweke, (1991))\nocite{geweke1991evaluating} makes to situation numerically solvable.

\vspace{5mm}
\noindent
In the one-dimensional case, $\sigma^2$'s prior comes from the Inverse-Gamma distribution. The uninformative prior is $f(\sigma^2) \propto \frac{1}{\sigma^2}$ (by the change of variable rule this corresponds to an uniform prior on $\textrm{ln}(\sigma^2)$). The multivariate version of the Gamma distribution is the Wishart distribution. The prior of inverse of $\boldsymbol{\varSigma}=\boldsymbol{\sigma \sigma}'$ is
\begin{equation}
f(\boldsymbol{\varSigma}^{-1}) \sim W_2(\boldsymbol{\Psi_\varSigma},\nu_{\boldsymbol{\varSigma}})
\end{equation}
By letting the hyperparameters of the inverse Wishart prior go to zero, we remain uninformative or diffuse on $\boldsymbol{\varSigma}$. Theoretically the degrees of freedom should be larger than or equal to the dimension of the matrix to ensure the draw to be invertible conditional that the hyperparameter $\boldsymbol{\Psi_{\varSigma}}$ is invertible. Thus the smallest number would be 2. Although we only draw from the posterior distribution, to be safe regarding the invertibleness we set $\nu_{\boldsymbol{\varSigma}} =3$. The degrees of freedom can be interpreted as the prior sample size (Gelman and Hill (2007))\nocite{gelman2007data} or the weight the prior mean gets compared to the data. The covariance matrix of the covariance matrix $\boldsymbol{\Psi_{\varSigma}}$ is set to standard deviations of $0.01$ and the correlation to $0.95$. The conditions based on inequality \eqref{restriction_kappa}, that is $\tilde{\kappa}$ to remain non-negative, are included in the model at this stage.  

\vspace{5mm}
\noindent
The conditional posterior distributions and the derivations can be found in Appendix F. The posterior hyperparameters are all functions dependent on the other parameters that is being conditioned on.

\vspace{5mm}
\noindent
Procedure
\begin{enumerate}
\item Draw long-term mean of zero-rates
\begin{enumerate}
\item If conditional posterior mean implies a low acceptance probability use exponential rejection sampling
\item Accept if the mean of the short rates under both measures and mean of the zero-rates implied by this draw is positive
\end{enumerate}
\item Draw the mean-reversion parameter of the zero-rates
\begin{enumerate}
\item If conditional posterior mean implies a low acceptance probability use exponential rejection sampling
\item Accept if positive
\end{enumerate}
\item Draw covariance matrix
\begin{enumerate}
\item Accept if it implies a positive mean-reversion parameter of the short rates
\end{enumerate}
\end{enumerate}

\section{Empirical application}
In this section we apply the Bayes algorithm to the data described earlier for $1,000,000$ simulations.
The output table is based on the noise decomposition of the covariance matrix. The average over all draws is shown, whereafter the 95\% Highest Posterior Density region (HPD95) and the 95\% Credible Interval (CI95) are reported, plus the standard deviation in the last column. The parameterization of $\boldsymbol{\varSigma}$ does not cause a great distinction between the two decompositions, therefore we do not show the outcomes based on the correlation decomposition.

\newpage
\noindent
\begin{table}[!h] 
\caption{Parameters noise decomposition} 
\begin{center} 
  \begin{tabular}{ l l l l l l l} 
    \hline 
& Average & HPD95 lb & HPD95 ub &CI95 lb & CI95 ub & St. Dev.\\ 
\hline 
$\kappa$ &   0.1647 &    2.7854$\times 10^{-5}$ &   0.3177 &    0.020404 &   0.3521 &   0.0862 \\ 
 $\tilde{\kappa}$ &   0.0205 &    1.6054$\times 10^{-3}$ &   0.0383 &    3.3314$\times 10^{-3}$ &   0.0407 &   0.0096\\ 
 $\mu$ &   0.0087 &     -7.3514$\times 10^{-3}$ &   0.0259 &   -5.8714$\times 10^{-3}$ &   0.0280  &   0.0080\\ 
 $\tilde{\mu}$ &   0.2224   &    1.0174$\times 10^{-3}$ &   0.4630 &   0.04430 &   0.6904 &   1.2155\\ 
 ${\theta}$ &   -5.575 &     -0.3382 &   0.2261 &   -1.1327 &   0.1539 &   328.2 \\ 
 $\Lambda_0$ &   -0.1611 &  -0.7297  &   0.4984 &  -0.6624 &   0.6002 &   0.2972\\ 
 $\Lambda_1$ & -20.9735 & -44.7955  &   2.3546 & -48.8694 &   0.1319 &  12.7564 \\ 
 $\sigma^2$ &    4.8574$\times 10^{-5}$ &    3.3154$\times 10^{-5}$  &    6.5454$\times 10^{-5}$ &    3.4334$\times 10^{-5}$ &    6.7204$\times 10^{-5}$ &    8.4214$\times 10^{-6}$ \\ 
 $\eta$ &    1.0854$\times 10^{-5}$ &    8.4224$\times 10^{-6}$ &    1.3444$\times 10^{-5}$ &    8.5954$\times 10^{-6}$ &    1.3684$\times 10^{-5}$ &    1.3004$\times 10^{-6}$ \\ 
    \hline 
\end{tabular} 
\end{center} 
\caption *{\small{\textit{The average over all $1,000,000$ draws for  $\tau_1 = 5$, $\tau_2 =20$, where $\tilde{\kappa}$ is solved numerically. The second and the third column show the lower- and upperbound of the $95\%$ highest posterior density region whereas the lower- and upperbound of the $95\%$ credible region are displayed in the fourth and fifth column respectively. The last column is the standard deviation based on draws.}}}
\label{table:noise_output}
\end{table} 

\begin{figure}[!h]
\centering
\caption*{Densities}
\begin{subfigure}[b]{0.45\textwidth} 
	\centering 
	\includegraphics[width=\textwidth]{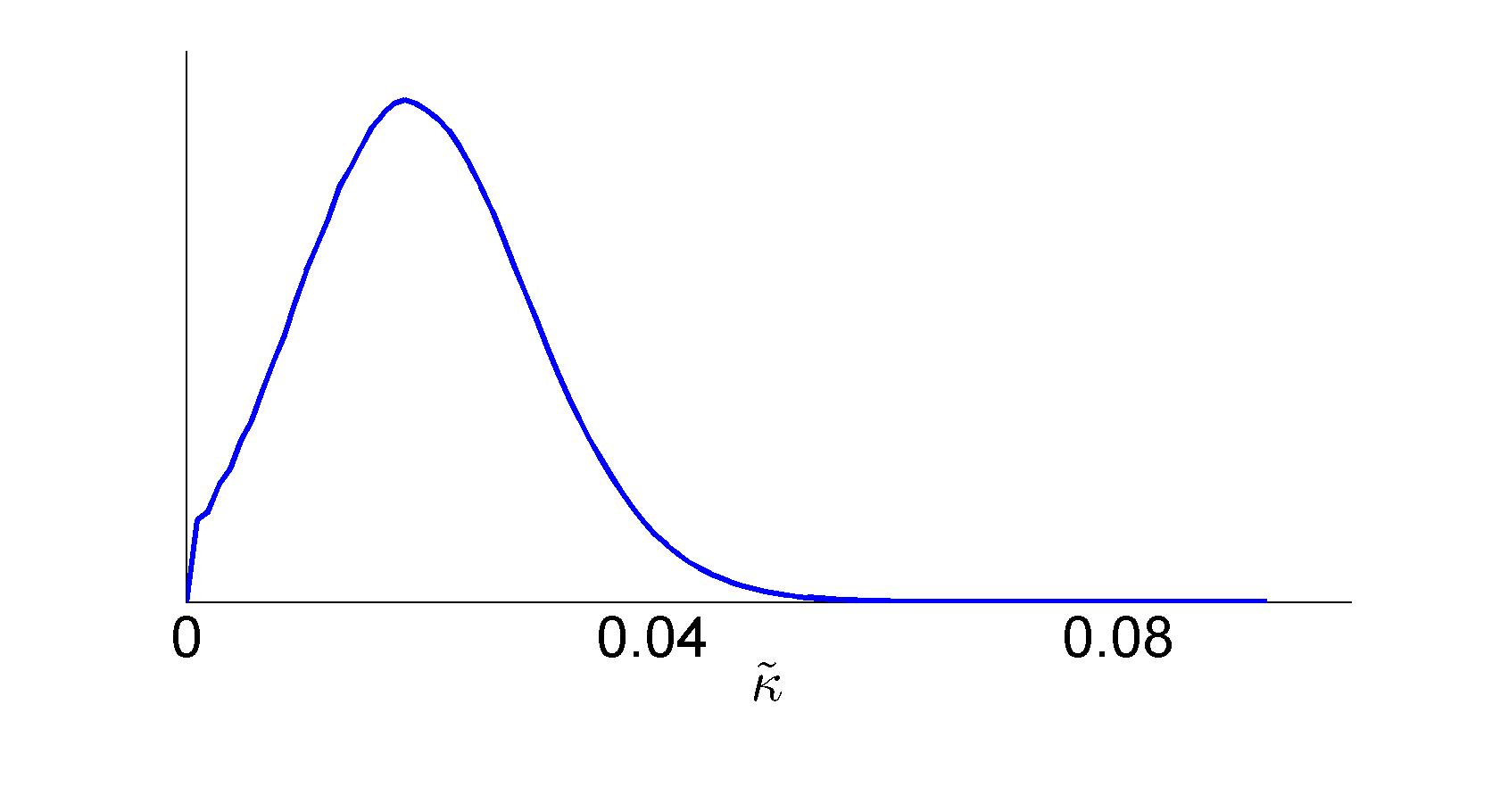}
	\caption{} 
\end{subfigure}
~
\begin{subfigure}[b]{0.45\textwidth}
	\centering
	\includegraphics[width=\textwidth]{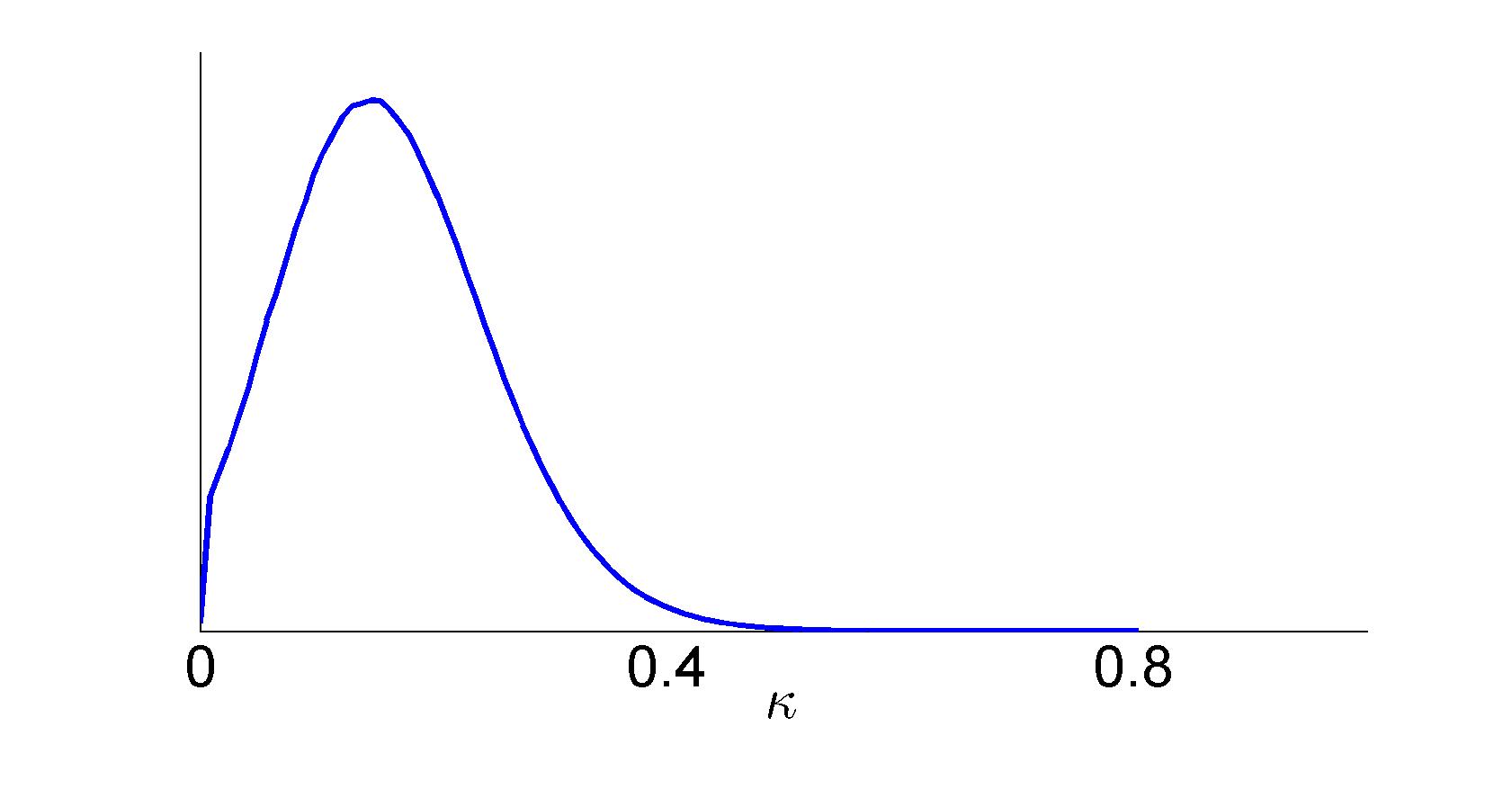}
	\caption{}
\end{subfigure}

\caption{{\small \textit{Densities for $\tau_1=5$ and $\tau_2=20$. The plots of $\tilde{\mu}$ and $\theta$ are adjusted to a visible mass density since the complete data set of the two parameters are extremely wide due to the large uncertainty.}}}
\label{fig:densities_noise1}
\end{figure}

\newpage
\begin{figure}[!h]
\centering
\caption*{Densities}
\begin{subfigure}[b]{0.45\textwidth}
	\centering
	\includegraphics[width=\textwidth]{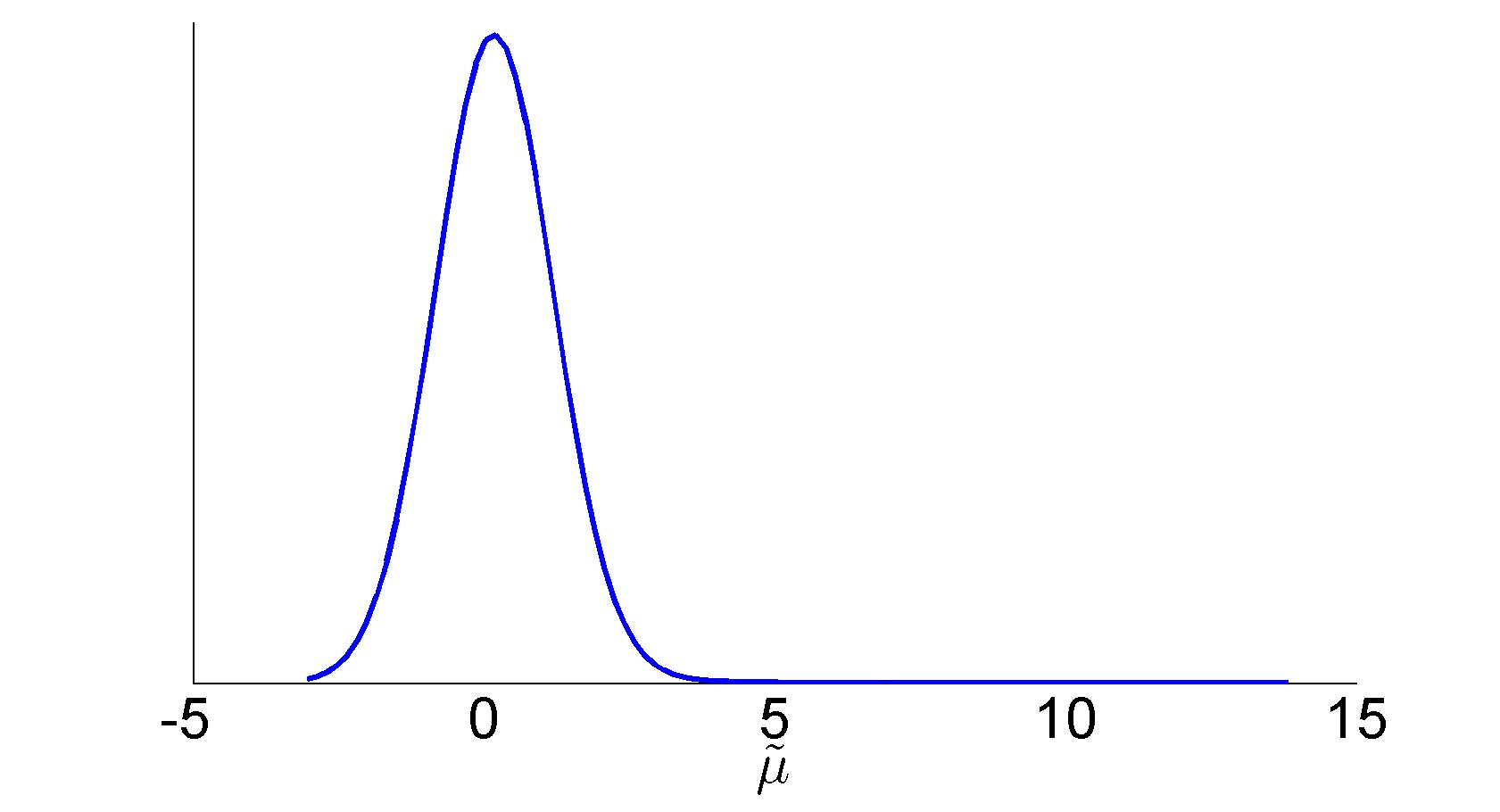}
	\caption{}
\end{subfigure}
~
\begin{subfigure}[b]{0.45\textwidth}
	\centering
	\includegraphics[width=\textwidth]{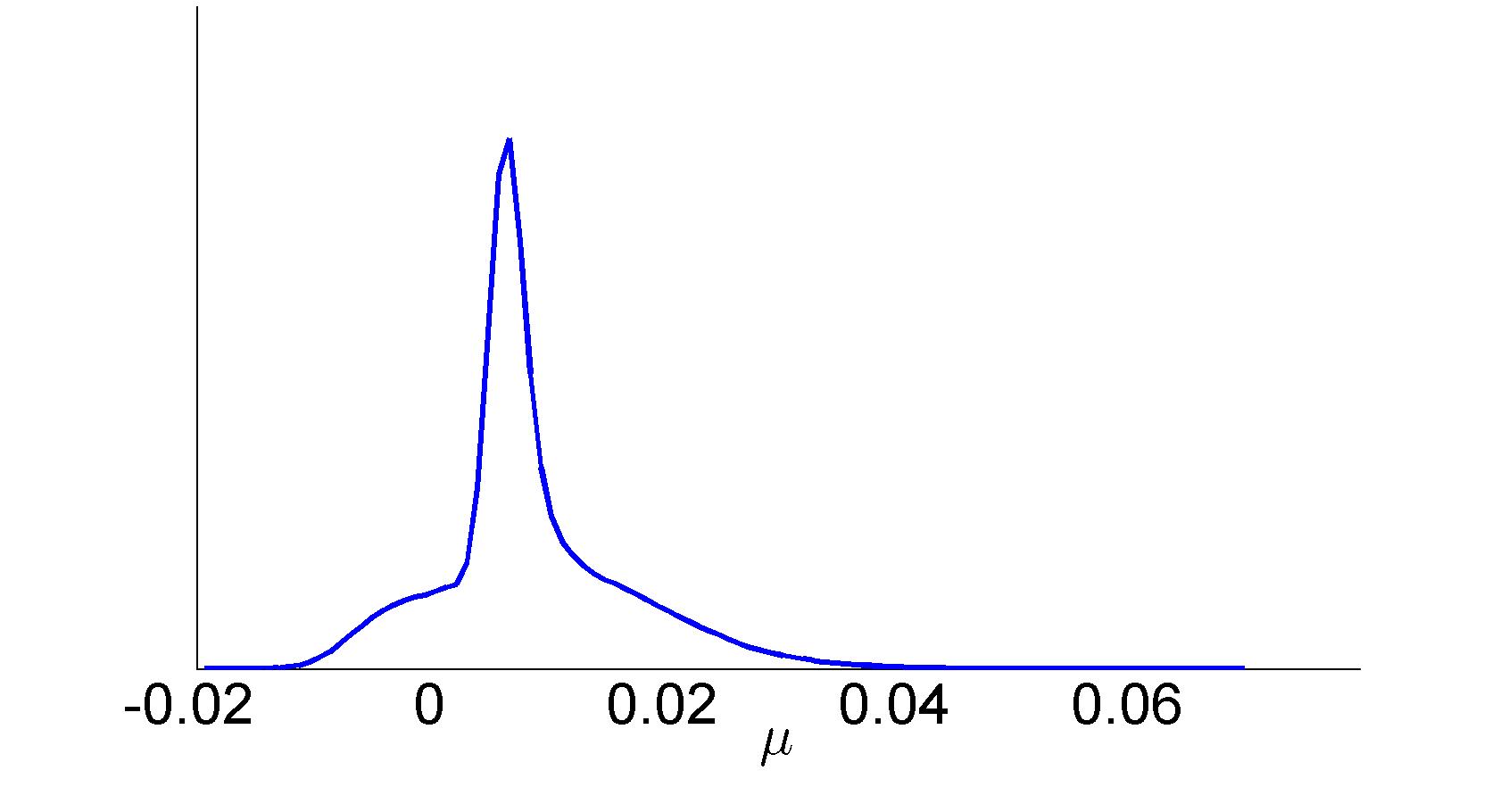}
	\caption{}
\end{subfigure}
~
\begin{subfigure}[b]{0.45\textwidth}
	\centering
	\includegraphics[width=\textwidth]{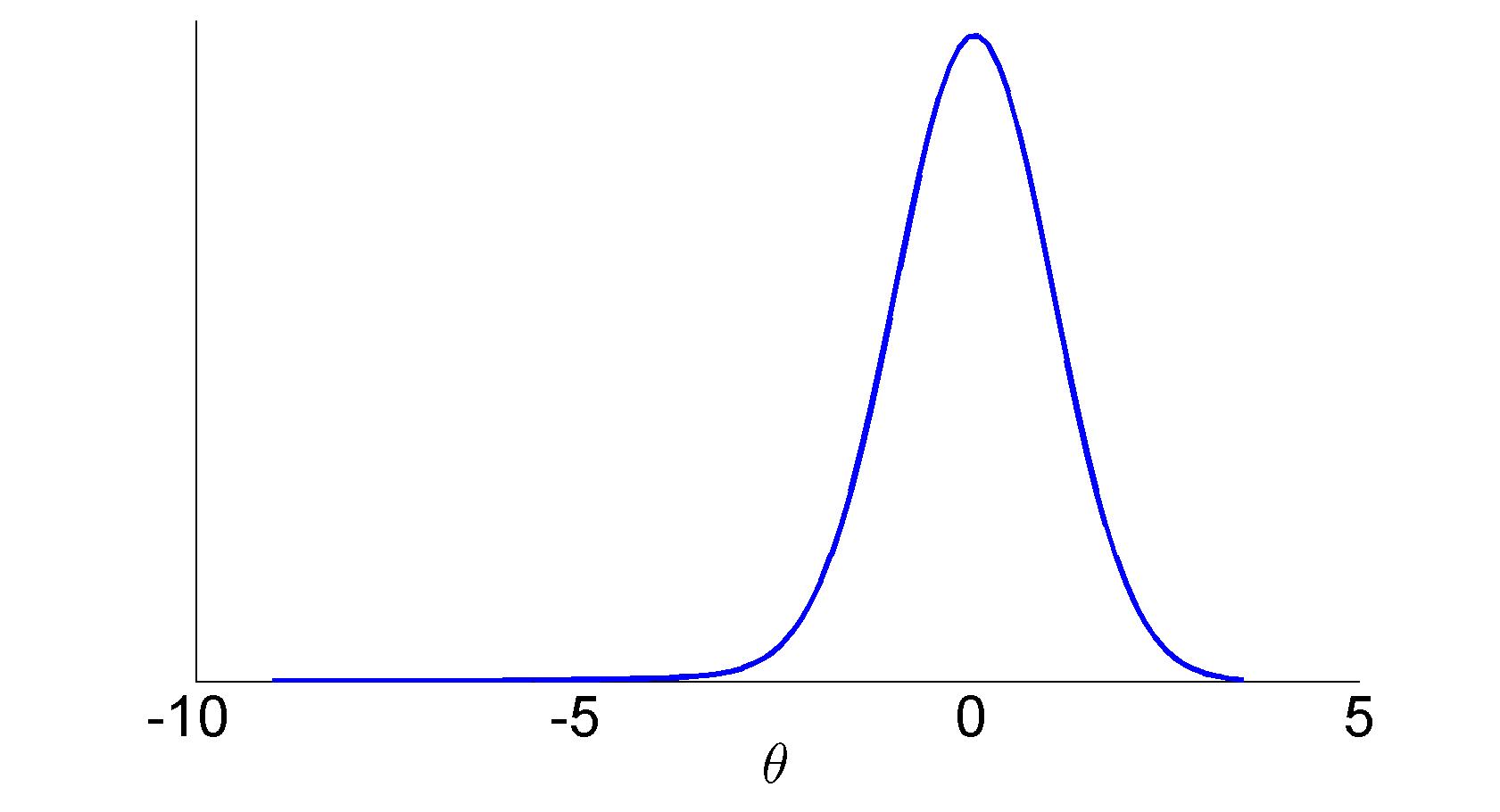}
	\caption{}
	\label{fig:theta_small}
\end{subfigure}
~
\begin{subfigure}[b]{0.45\textwidth} 
	\centering 
	\includegraphics[width=\textwidth]{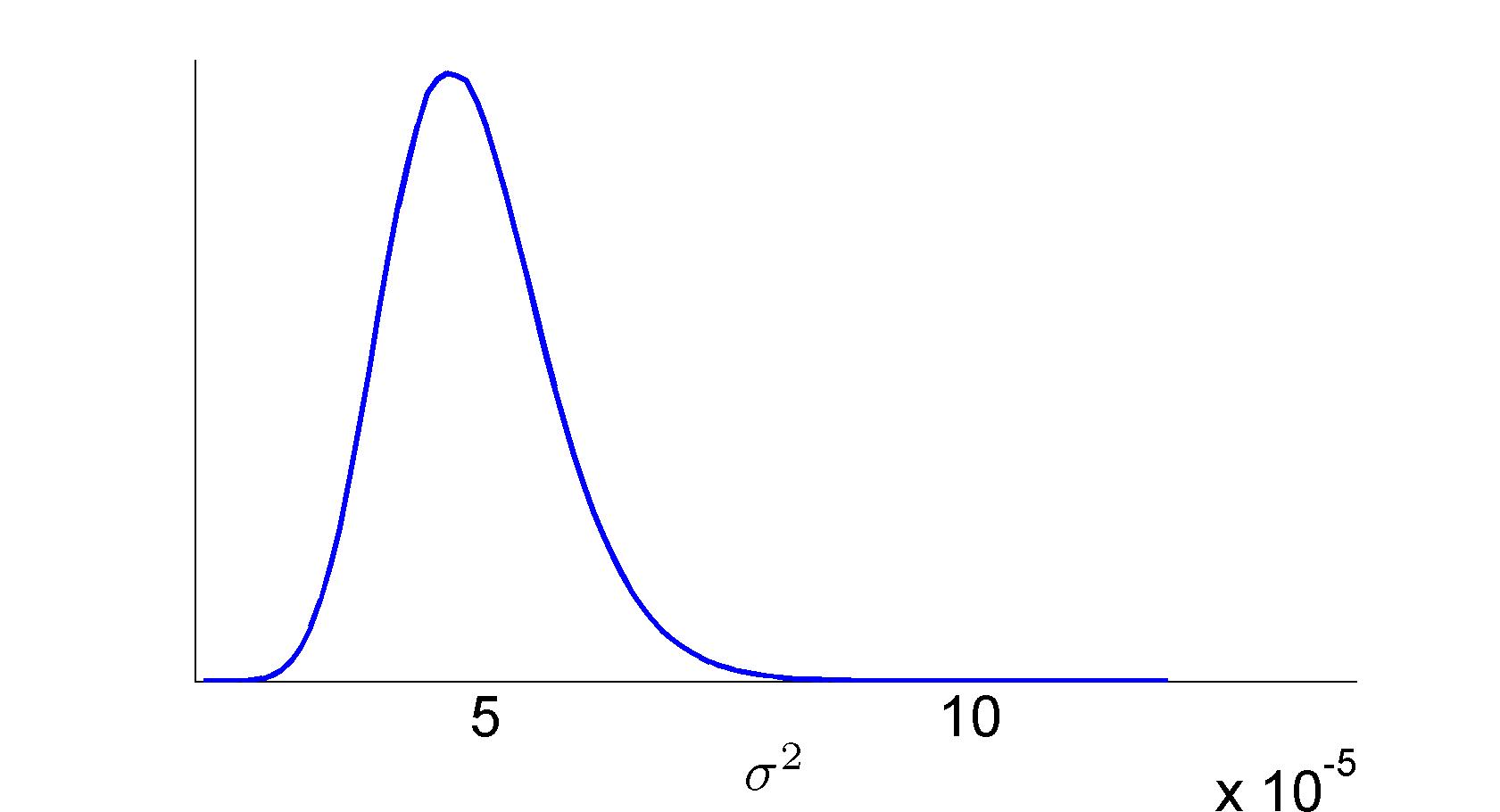}
	\caption{} 
\end{subfigure}
~
\begin{subfigure}[b]{0.45\textwidth}
	\centering
	\includegraphics[width=\textwidth]{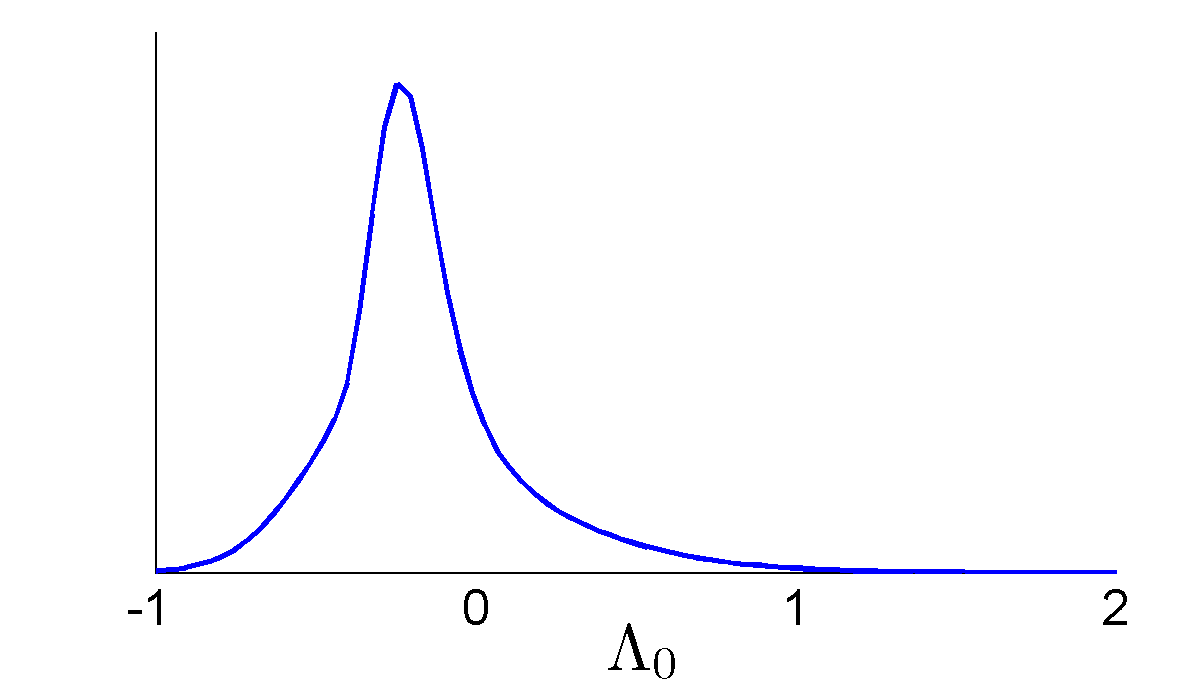}
	\caption{}
\end{subfigure}
~
\begin{subfigure}[b]{0.45\textwidth}
	\centering
	\includegraphics[width=\textwidth]{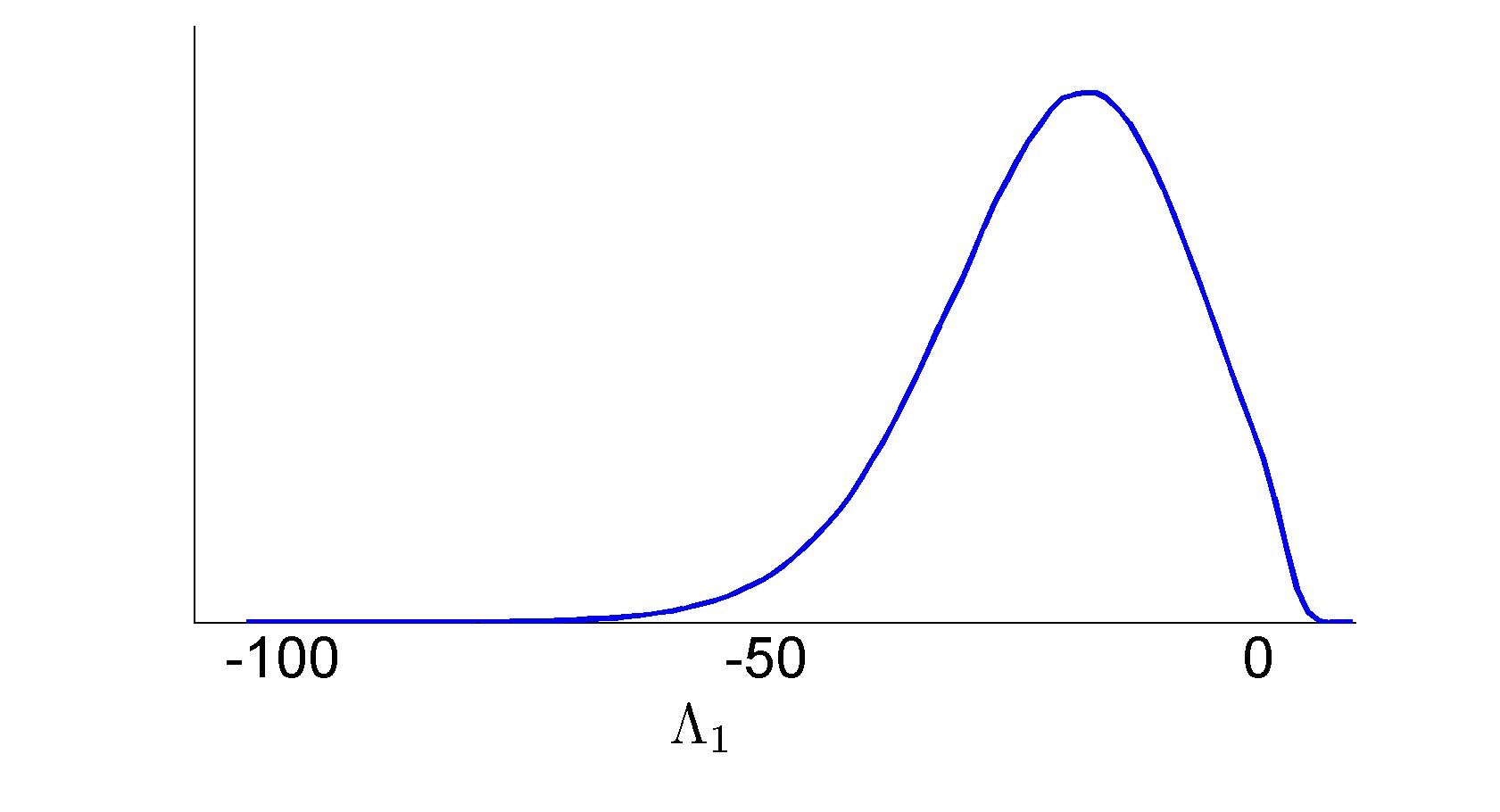}
	\caption{}
\end{subfigure}
~
\begin{subfigure}[b]{0.45\textwidth}
	\centering
	\includegraphics[width=\textwidth]{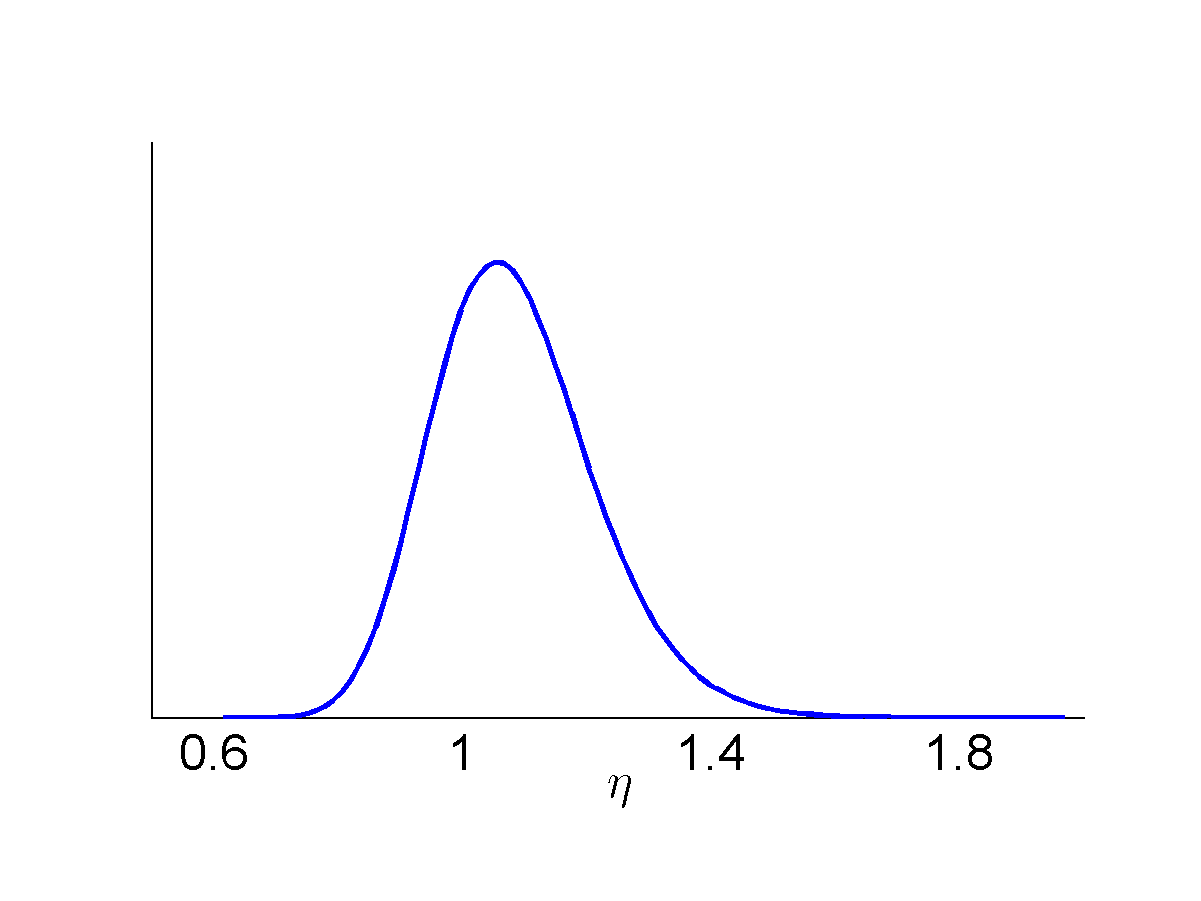}
	\caption{}
\end{subfigure}
\caption{{\small \textit{Figure \ref{fig:densities_noise1} continued. }}}
\label{fig:densities_noise2}
\end{figure}

\newpage
\noindent
The density of $\theta$ is hard to identify since the range is extremely wide. If we plot the figure without the 10,000 smallest draws the mass of the body can be observed in Figure \ref{fig:theta_small}. This also explains the relatively large standard deviation compared with the frequentist approach, however the HPD and CI are much smaller. The conclusion that can be drawn is that the uncertainty of the long-term mean is very large. Hence it is not reliable to trust point estimates with lengthly maturities. The mass of $\theta$ lies in a reasonable range, but due to some exceptional outliers the average and the standard deviations are so negative and large respectively. The cause of these outliers is the unit root problem. From the cMLE this difference cannot be seen since the two outputs are the estimation and the standard deviation. The density of the Bayesian output shows the non-normal shape.

\begin{figure}[h!]
\centering
\caption*{Sensitivity}
\begin{subfigure}[b]{0.45\textwidth} 
	\centering
	\includegraphics[width=\textwidth]{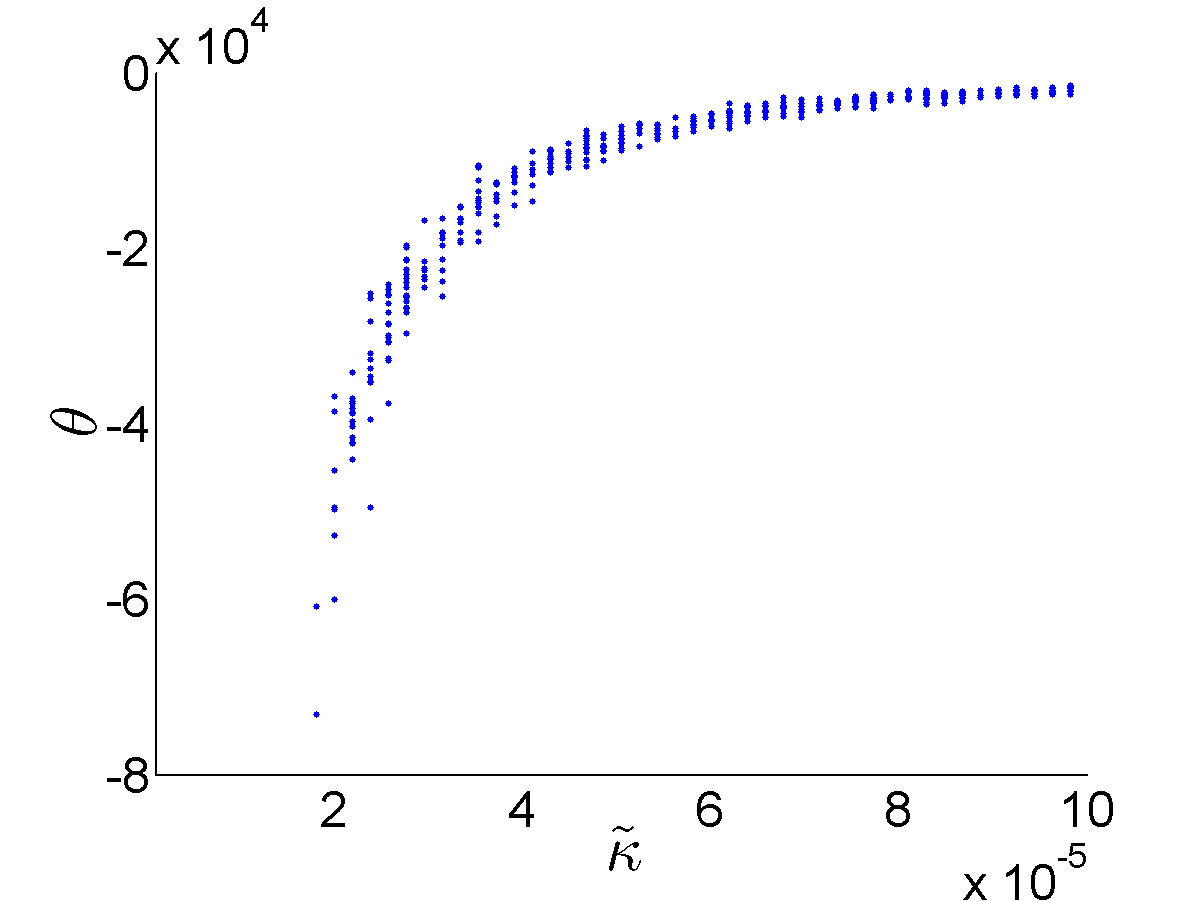}
	\caption{Zoom low values $\tilde{\kappa}$} 
\end{subfigure}
~
\begin{subfigure}[b]{0.45\textwidth} 
	\centering
	\includegraphics[width=\textwidth]{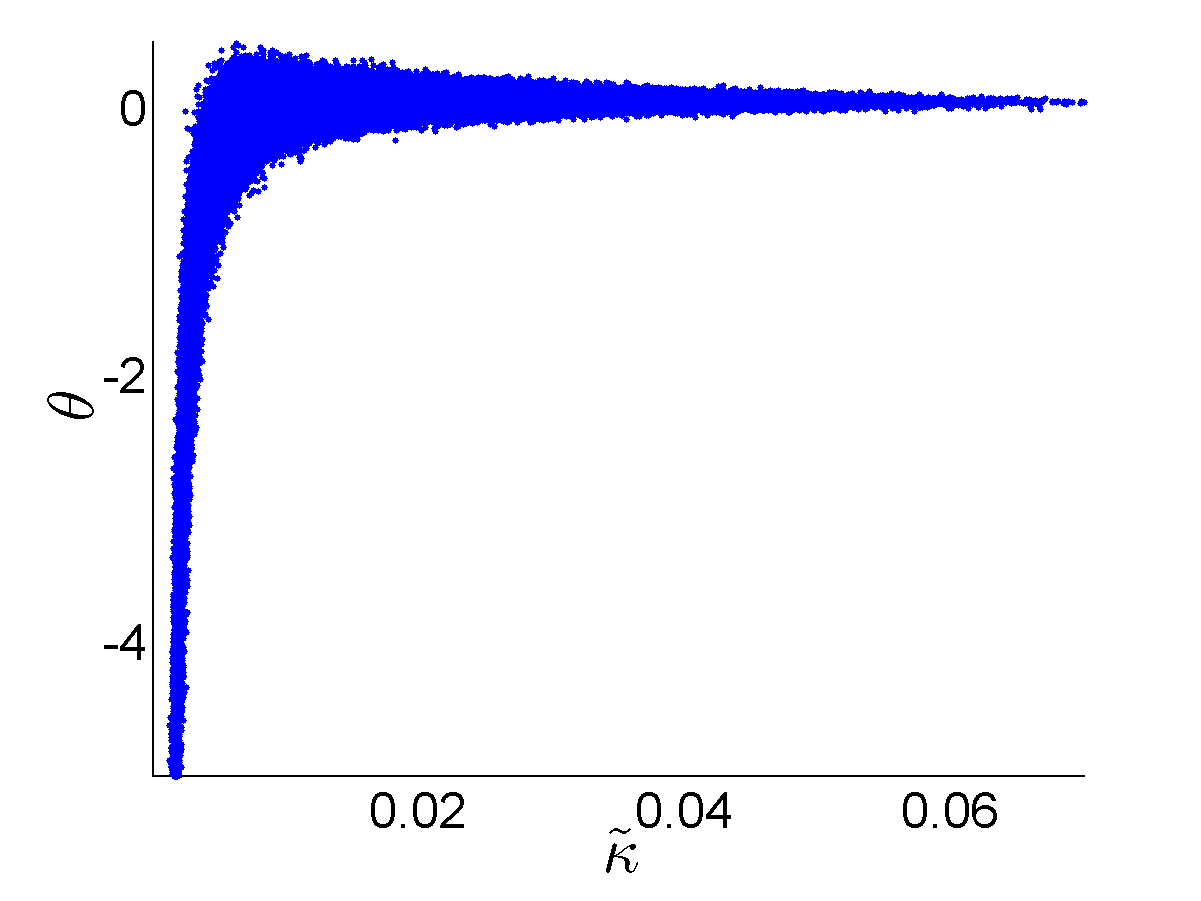}
	\caption{Zoom large values $\tilde{\kappa}$} 
\end{subfigure}	
\caption{\small {\textit{Scatter plot of the draws $\theta^{(i)}$ against $\tilde{\kappa}^{(i)}$ based on the data set described and maturities $\tau_1 = 5$ and $\tau_2 = 20$.}}}
\label{scatter_plot}
\end{figure}

\vspace{5mm}
\noindent
The mean-reversion going to zero, means that the lagged rate in the process goes to one, $(1-\tilde{\kappa} h) \rightarrow 1$ for $\tilde{\kappa} \rightarrow 0$, raising the problem of a unit root. The unit root problem in a general AR process implies that the scalar in front of the lagged variable going to one causes the variance of the process to go to infinity. When $\tilde{\kappa}$ is close to zero, ${\theta}$ is very uncertain according to its wide interval, which means that when the mean-reversion is very slow the model does not know to which level it converges. Since the convergence rate is so low the time period untill the ultimate level goes to infinity, and therefore the uncertainty about the long-term mean has no effect. While if the convergence rate increases it becomes more and more apparent to which long-term mean will converge and since it moves quicker towards this level the importance of knowing this level has also increased. This pattern can be recognised Figure \ref{scatter_plot}. The scatter plot $(\tilde{\kappa},\theta)$ shows that for small values of $\tilde{\kappa}$, the uncertainty of the ultimate level is characterised by a wide spread of $\theta^{(i)}$. Actually, $\theta$ is largely determined by $\tilde{\kappa}$, if the cross-sectional mean reversion goes to zero, the ultimate level goes to minus infinity. To better see the dependence we split the graph in low values for $\tilde{\kappa}$ and a wide axis for $\theta$ and a graph for the relatively high values of $\tilde{\kappa}$. Hence we are interested the parameter $\tilde{\kappa}$ and we want to measure the uncertainty of the mean-reversion since this is the factor that determines the extrapolations.

\noindent

\vspace{5mm}
\noindent
The two $95\%$ Bayesian intervals of $\tilde{\kappa}$ show to be in a range of $(0, 0.04)$. Whereas the cMLE approach was extremely uncertain about this rate. Especially all the Bayesian intervals show reasonable ranges for the parameters of interest, which also points out the difference with a single estimated standard error. The impact of the variance and hardly tracable effect of the variance of the separate parameters on the extrapolation will be showed in the next section.
The mean-reversion is around 0.02, as expected lower than under the physical measure. As Bauer (2011)\nocite{bauer2011term} states that if one believes in the absence of arbitrage then both probability measures' parameters should be close to each other which confirms his finding of favorable models restricting $\Lambda_1$ going to zero. "Because typically many cross-sectional observations are available the Q-dynamics can be precisely estimated". Supporting this statement, we found a much smaller standard deviation for $\tilde{\kappa}$ than under the historical measure. However, the restriction on $\Lambda_1$ equal to $0$ is just on the edge of the credible interval and therefore putting up this condition upfront asks for a very informative prior. Note that the analytical VAR(1) model does not include a direct prior on the dependence between the two measure as we assumed no correlatin upfront.    

\begin{figure}
\centering
\caption*{Convergence}
\begin{subfigure}[b]{0.45\textwidth} 
	\centering 
	\includegraphics[width=\textwidth]{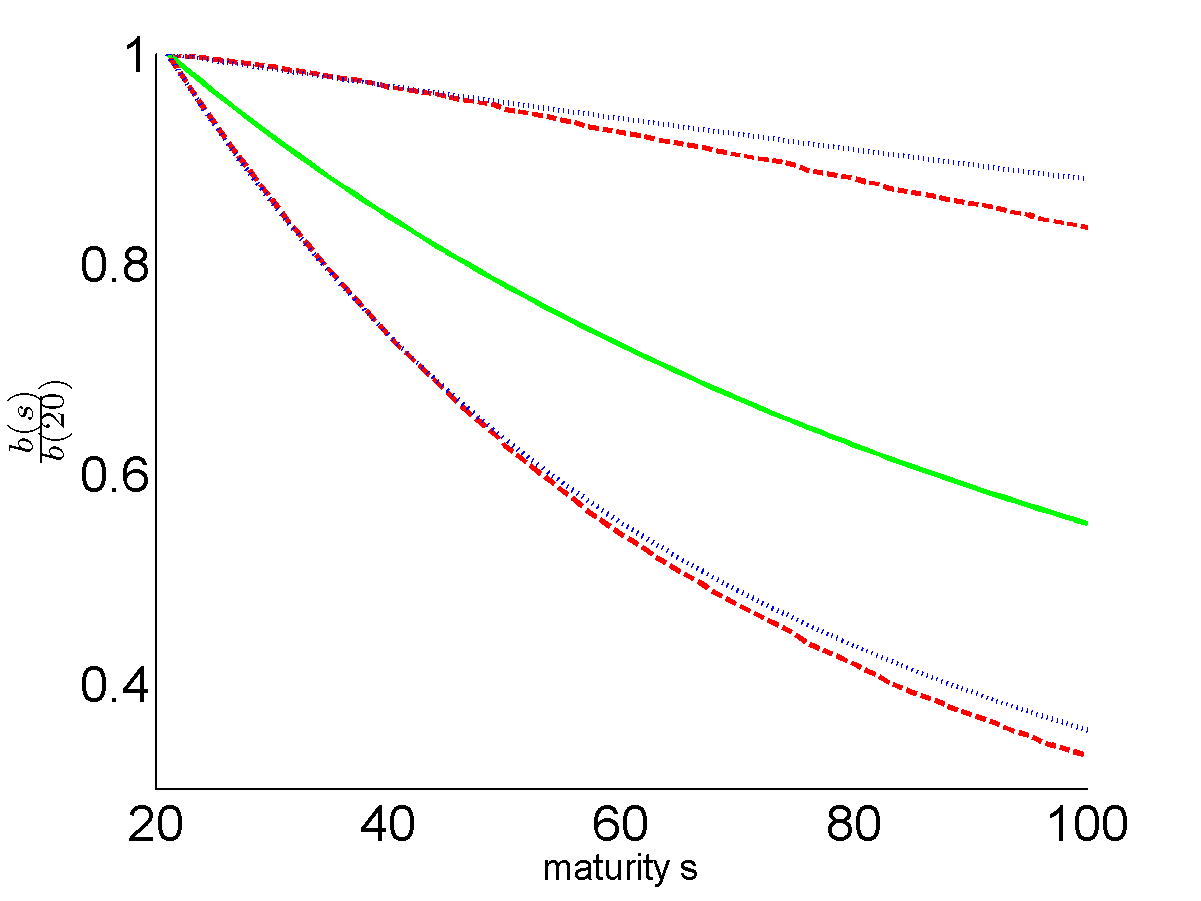}
	\caption{Convergence speed over time} 
\end{subfigure}
~
\begin{subfigure}[b]{0.45\textwidth} 
	\centering 
	\includegraphics[width=\textwidth]{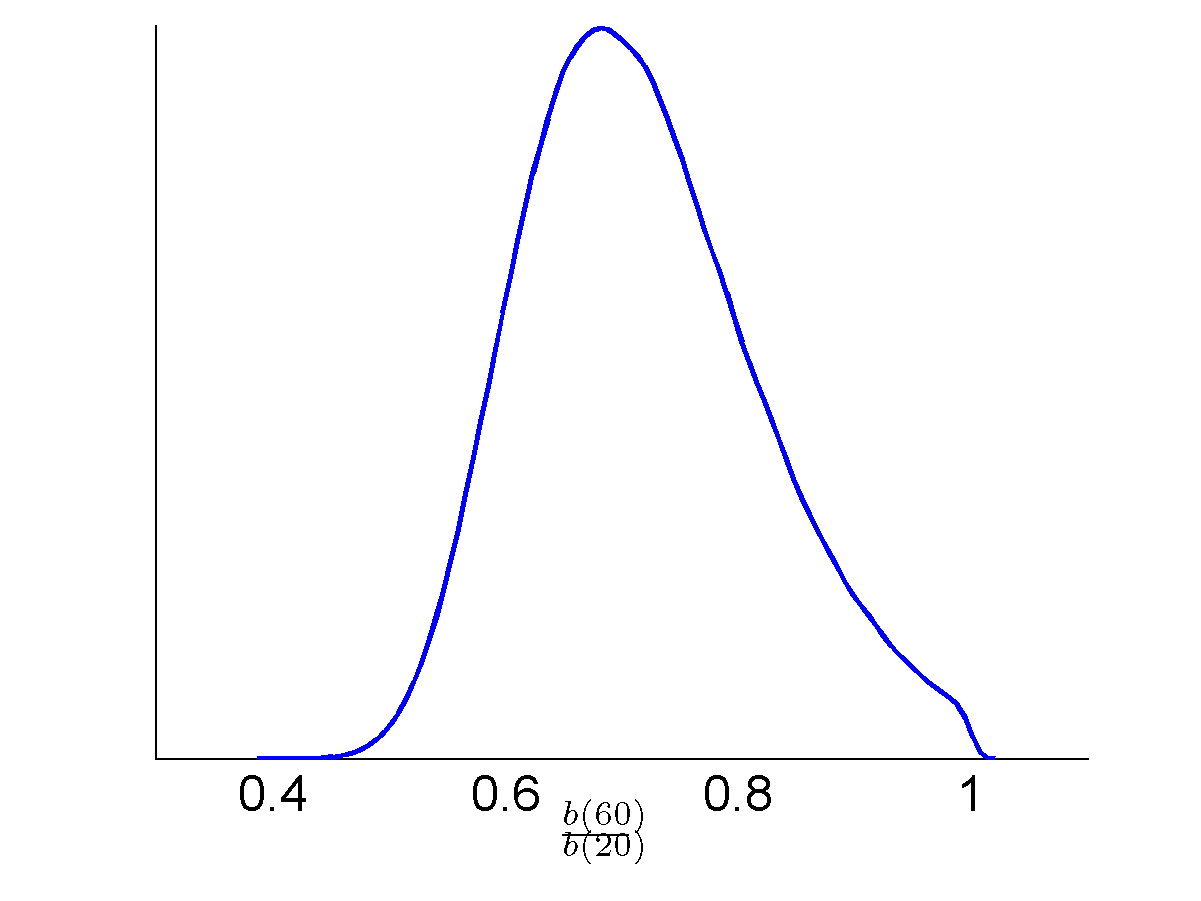}
	\caption{Density} 
\end{subfigure}
\caption{\small {\textit{On the left the average over all different $\tilde{\kappa}$s where $\tau=20$ is the last liquid point and $s \in (\tau+1, \tau+2,...)=(21,22,...,100)$. The red dashed line represents the $95\%$ highest posterior density region wheras the blue line the $95\%$ credible interval is, the green line is the average. On the right the density for a fixed extrapolation point $s=60$ is shown based on all simulations of $\tilde{\kappa}$.}}}
\label{fig:bs_btau}
\end{figure}

\vspace{5mm}
\noindent
The convergence speed between an unknown rate on a longer horizon and a known rate on a closer horizon comes from the relation \eqref{y_t}.
We only need the mean-reversion parameter for $\frac{b(s)}{b(\tau)}$. The interpretation of this ratio can be expressed in terms of volatility as well as convergence speed. Firstly, the ratio is the relative volatility of a $s$-year maturity rate to a $\tau$-year maturity rate since $\sqrt{\textrm{var}}[y(\tau)]=\textrm{vol}[y(\tau)]=b(\tau)\sigma$. Hence $\frac{b(s)}{b(\tau)}=\frac{\textrm{vol}[y(s)]}{\textrm{vol}[y(\tau)]}$. Secondly, it shows the speed how fast it moves towards the long-term mean. The ratio behavior is depicted in Figure \ref{fig:bs_btau}. Trivially tomorrows rate depends heavily on todays rate. This relation declines over time which can be seen in the figure on the left. On average the ratio is 0.7 between the forecasted 60-year maturity and the last liquid point of 20 year, this indicates that the relation between these two maturities is still there, contrary to the idea linked to the UFR and Smith-Wilson methodology, of a constant ultimate level at 60 years.

\vspace{5mm}
\noindent
The correlation $\rho$ between the 5 year and 20 year rates is 0.77. Neither too high to still catch the curvature, nor too low, which can be more easily interpreted by the error term $\eta$. An average noise term of $1.1 \times 10^{-5}$ indicates that the model based on these two maturities does not cause too much noise. As already pointed out, the input choice is compared with standard curvature research relatively long termed. This supports the one factorisation and fits the aim of the method that is interested in extremely far dated rates.

\vspace{5mm}
\noindent
The continuous relation between the $N$-maturity one-year forward rate $f_t^{(N \rightarrow N+1)}$ and an observed zero-rate (Cochrane (2001))\nocite{cochrane2001asset} has the following limiting value.
If the maturity goes to infinity, the Ultimate Forward Rate (UFR) becomes
\begin{eqnarray}
\lim\limits_{N\rightarrow \infty} f_t^{(N \rightarrow N+1)} &=& \theta \nonumber
\end{eqnarray}
The mean-reversion parameter between the zero-rate observed with maturity $\tau$ and the $N$-year forward rate is
\begin{equation}
\tau e^{-\tilde{\kappa} N}\frac{(1-e^{-\tilde{\kappa}})}{(1-e^{-\tilde{\kappa} \tau})} \nonumber
\end{equation}
obtained by simply rewriting the expression for $z(\tau)$ in term of forward rates.
Concerning the recent debate about the UFR, $\tau$ is set to the last liquid point and extrapolation period. Common choices by Dutch pension funds following the rules of Solvency II were a last liquid point of $20$ and the moment of reaching the UFR at $60$ years. With $\tilde{\kappa}=0.02$ the mean-reversion rate between  $y(20)$ and $f_t^{(60 \rightarrow 61)}$ is about $36\%$. Similar to what we just discussed about the relation between the two zero-rates, the dependence between the zero-rate and the forward rate diminishes if the extrapolated forward rate moves further away. This general tendency is in line with the modelled UFR technology, although there still dependency left after 60 years. 

\noindent



\section{Extrapolation}
Common extrapolation methods are the Nelson-Siegel method and the Smith-Wilson method. The Nelson-Siegel (NS) function extrapolates the long end of the yield curve based on a single set of shorter maturity rates. The long end behaves rather constant which is a feature appreciated by practitioners. However, the extrapolation of today is different than the extrapolation of tomorrow, therefore the volatility at the long end is high as for every cross section a different curve is obtained. Note also that the ultimate level is highly dependent on the last observed rate and thus unexpected shifts cause high uncertainty towards the long end. The high variability of this technique rises the incentive for a model that moves cross sectionally to a stationary rate at the very long end. The Smith-Wilson (SW) method is an interpolation method that fully uses the idea of an ultimate constant level. As an interpretation of the models, we can rank the models from volatile to constant by NS, Vasicek and SW respectively. What we like to measure is the uncertainty of the long-term rate. Thus whether the data shows a constant level for very long maturities or high volatile extrapolations.

\subsection{UFR extrapolation}
For pension funds and insurance companies recent developments about pricing of long-term obligations is under debate. In some countries the UFR is applied by central banks as explained in Solvency II. We apply the Smith-Wilson\footnote{Fitting Yield curves with long Term Constraints, Smith, A. and Wilson, T. Research Notes, Bacon and Woodrow, 2001.} smoothing technique (Thomas and Maré (2007)\nocite{thomas2007long} and some implementational notes from the Norway (2010)\footnote{A Technical Note on the Smith Wilson Method, The Financial Supervisory Authority of Norway, 2010.}) to the swap curve from September 2013 with an UFR of $4.2 \%$, a last liquid point of 20 years and the aim of reaching the UFR in $60$ years from now by approaching it by a deviation of at most 3 basis points. The graph shows the curvature based on these input choices, plus the swap curve consisting of the few quoted rates for longer maturities. Since it is an ongoing topic a recent report by the Dutch UFR committee (October 2013) suggested an ultimate level that is more historically founded and a smoothing technique between the market data still available beyond the last liquid point which is taking into account by a decreasing weight based on the Vasicek model. By drawing the point estimates for all maturities the strength of this technique can only be assessed by the addition of a measure of uncertainty, such as variances.

\noindent
\begin{figure}[!h]
\label{fig:SW}
\centering
\caption*{Smith-Wilson} 
\begin{subfigure}[b]{1\textwidth}  
	\includegraphics[width=\textwidth]{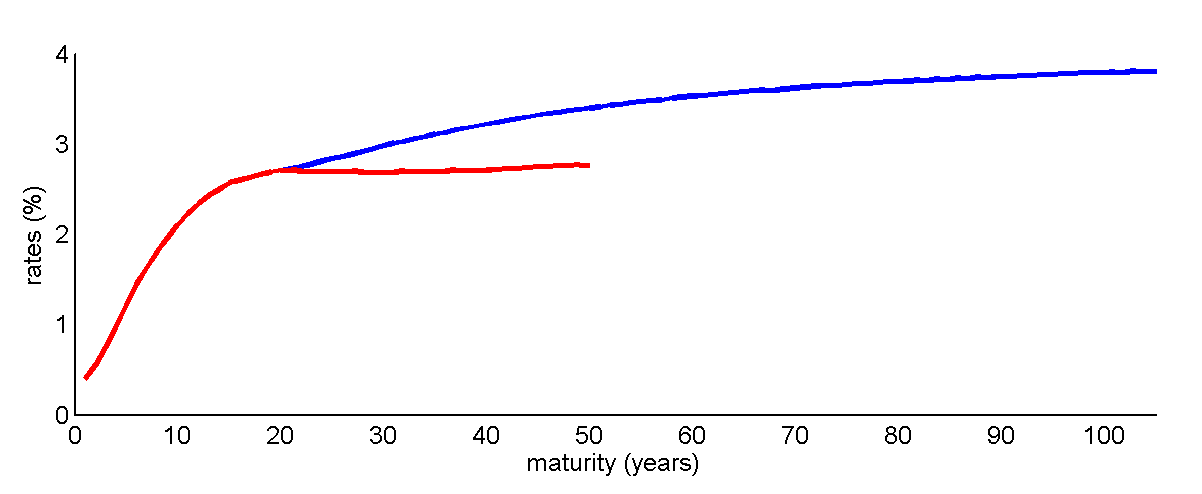}
\end{subfigure}
\caption{\small {\textit{The Smith-Wilson methodology applied to the zero curve from September 2013 where the last liquid point is $20$ years, the time of reaching the UFR is $40$ years later, thus at $60$ years from September 2013 and the UFR level is set at $4.2\%$, which is reached within $3$ basis points. The red line shows the original input and the blue line is the extrapolated curve by Smith-Wilson}.}}
\end{figure}

\noindent
By construction the long term yield is completely certain as it is chosen upfront. It is therefore highly questionable whether this reflects market consistency. In the previous section we saw that the dependence of the forward rate declines by increasing extrapolation time, however $f_t^{(60 \rightarrow 61)}$ including the uncertainty bounds is still far from independent. We know from the data that the mean reversion is low, indicating that the horizon of an ultimate level is extremely far and consecutive resulting in extreme ultimate levels, while $\tilde{\kappa}$ should be large in order to support the UFR methodology.

\subsection{Nelson-Siegel extrapolation}
For every time series data set a different cross sectional extrapolation is obtained via the Nelson-Siegel method (1987)\nocite{nelson1987parsimonious}. This technique fits the parameters by a single curve and extrapolates the curve based on these fitted parameters.
\begin{equation}
z(t) = \beta_0 + \beta_1 \frac{1-e^{-t/\tau}}{t/\tau} + \beta_2 \left(\frac{1-e^{-t/\tau}}{t/\tau} - e^{-t/\tau}\right)
\end{equation}

\begin{figure}[!h]
\centering
\caption*{Nelson-Siegel}
\begin{subfigure}[b]{1\textwidth} 
	\centering 
	\includegraphics[width=\textwidth]{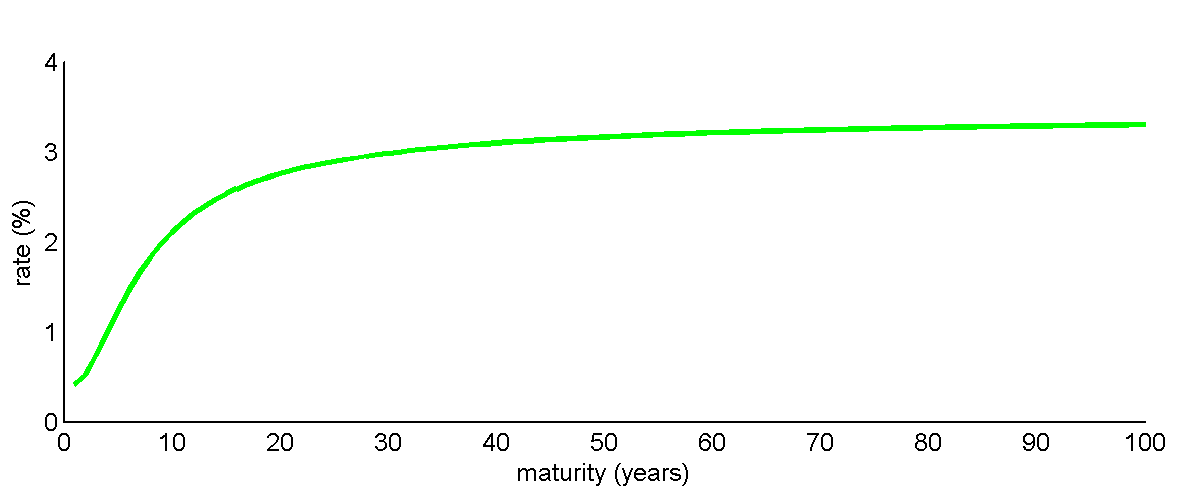}
\end{subfigure}
\caption{{\small \textit{The Nelson-Siegel function applied to the 1,2,3,...,20 year zero rates from September 2013 by the least square calibration. The estimates for the parameters are $\tau=1.93, \beta_0=0.03, \beta_1=-0.03, \beta_2=-0.04$. All yearly rates from 1 to 100 are calculated based on these estimates.}}}
\end{figure}

\noindent
We fitted the first 20 data points of the zero swap curve by least squares and extended the curve. The direction of extrapolation is rather flat compared with the Smith-Wilson method resulting in lower rates for long maturities than the UFR level. This is caused by relatively low market rates compared with the historic data set. Another characteristic of the Nelson-Siegel technique is that the extrapolations are highly volatily since every time a quoted price changes the complete extrapolation is affected by this. Especially movements in the last liquid rate causes large shifts in the ultimate level due to the straight extension.

\newpage
\subsection{Bayesian extrapolation}
Now we apply the described Bayesian approach, where we model the term structure by the affine Vasicek model under the assumption of parameter uncertainty. We select the cross sectional maturities of $5$ and $20$ years for the complete time spanned by the data. Furthermore, in the figure below we use the zero rates from September 2013 as last observed rates which are market consistent up to 20 years and then extend the curve by this papers' method. 

\noindent
\begin{figure}[!h]
\centering
\caption*{Bayesian extrapolation}
\begin{subfigure}[b]{\textwidth} 
	\centering 
	\includegraphics[width=\textwidth]{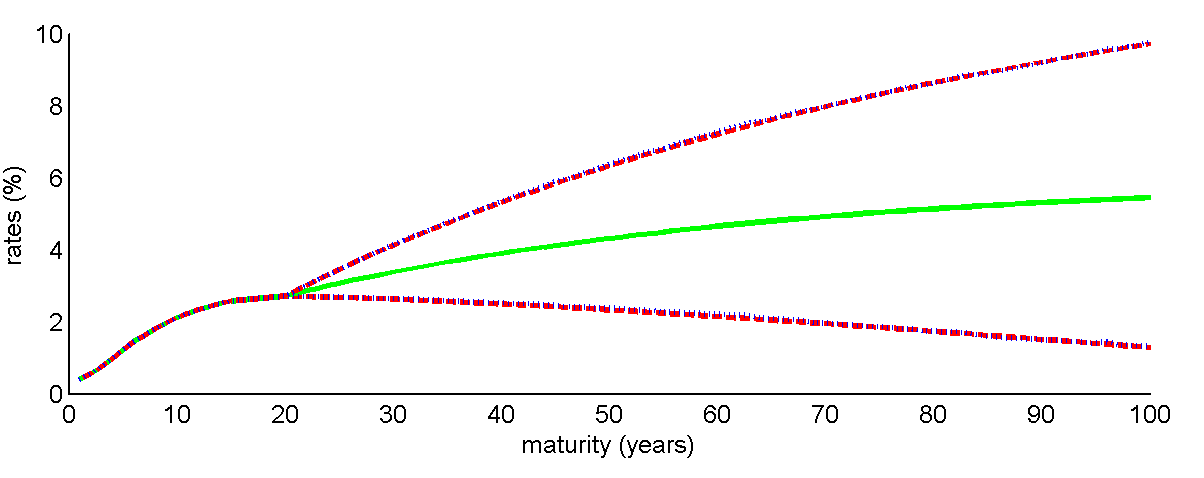}
	\caption{HPD and CI} 
\end{subfigure}
~
\begin{subfigure}[b]{1\textwidth}
	\centering
	\includegraphics[width=\textwidth]{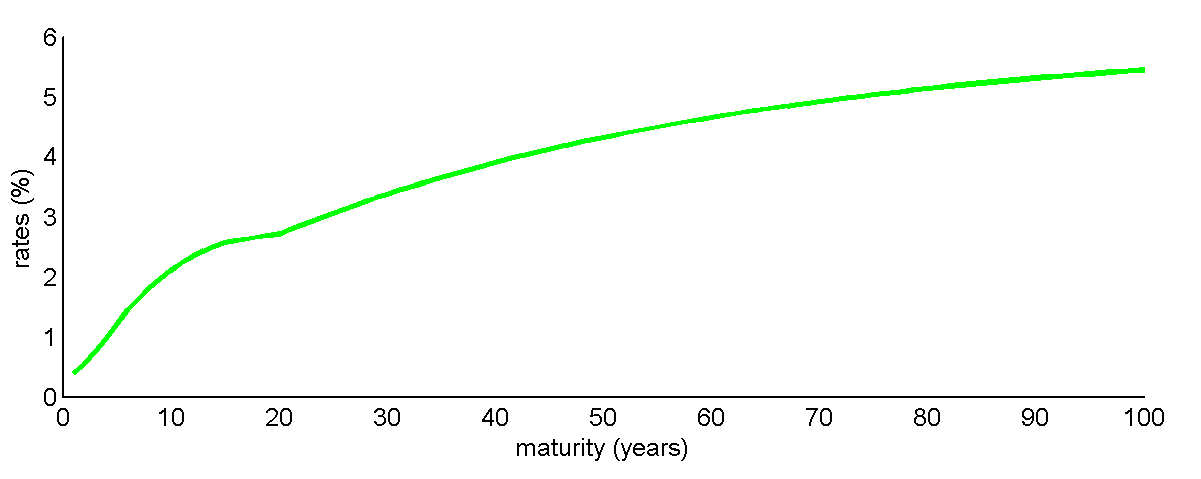}
	\caption{Average}
\end{subfigure}
\caption{\small {\textit{Figure (a) shows the average development, in green, based on the Bayesian draws plus the $95\%$ highest posterior region by the dased red line and the $95\%$ credible region by the dotted blue line. On the right the same curve without the confidence regions, thus a zoom of the average development in green. The curves are based on the 5- and 20-year maturity swap rates from 2002 till September 2013, whereas the first 20 maturities are the swap rates of September 2013 only. From the last dependend point of 20 year the extrapolation starts and is plotted untill the maturity of 100 years. The rates are represented in percentages.}}}
\end{figure}

\newpage
\noindent
From the extrapolation we can see that the point estimate has a higher slope and continues increasing after the point where the UFR level was kept constant. The strength of the methodology used in this paper is the addition of the HPD and CI under the positiveness restriction. The ranges show that the 100-year rate is in between 1\% and 10\% with $95\%$ confidence, a economical realistic range for interest rates but actually indicating a lack certain estimations. The UFR method and the Nelson-Siegel method both fall within the uncertainty sets. Compared with the other two methods, the fact that this model is solely focusing on the extremely long termed rates and thus only uses relatively long maturities as input is robust method and links the input and output consistently. While the NS approach uses relatively short rates in order to forecast rates up to a century.

\section{Robustness}
The discrete mean-reversion parameter under the physical measure in the short rate and zero-rate model are similar by construction. Therefore our discrete estimation can be compared to the discrete estimates of Chan, Karolyi, Longstaff, and Sanders (1992)\nocite{chan1992empirical} and A\"it-Sahalia (1996)\nocite{ait1996testing} of $\alpha_1 h$ which equals $-\kappa h$. CKLS estimate ranges from -0.18 to -0.59 for monthly observations of the one-month Treasury yield and based on daily observations of a one-week Eurodollar rate. A\"it-Sahalia's mean-reversion ranges from -0.014 to -0.038. For the single time series estimate of $\kappa$ we found 0.1604 hence comparable to $-0.1647\cdot \frac{1}{12} =-0.01373$, lying in a range from almost zero (slightly negative) to -0.02648 based on the HPD 95\% being a subset of both CKLS as A\"it-Sahalia's intervals.

\vspace{5mm}
\noindent
As a robustness check we applied the Bayesian procedure also to different choices of maturity sets of input and different hyperparameters of the priors. The sensitivity of the results for these choices shows to be small (see Appendix H for the sensitivy analysis). Also the ACF, CUSUM and Geweke tests show no convergence problem for the simulations (see Appendix I). 

\vspace{5mm}
\noindent
The continuous autoregressive gaussian affine model with parameter uncertainty is a theoretical model that can be generally applied to different data set and the model can be adjusted and extended if necessary. Here we applied the model as an illustrative example since there is no closed form solution for extrapolating with parameter uncertainty, but the solution is based on numerical procedures.

\section{Conclusion}
The ability to allow for parameter uncertainty in the Vasicek model under the condition that the mean and mean-reversion parameters are positive, makes the Bayesian setting attractive for interest rate modelling. We extrapolated the term structure of interest rates by the use of a data set consisting of rates with two different maturities. In this bivariate normal process the implied parameters are analytically solvable by the addition of a correlation or noise term. The conditional maximum likelihood estimators lead to broadth variances and negative means. The specification of parameter uncertainty in the affine zero-rate model resulted in realistic 95\%  credible intervals and highest posterior density regions. The range of extrapolation shows that the rate can be in between 1\% and 10\% based on an extrapolation from 20 till 100 year. Hence the uncertainty is so large that trusting a point estimate is not appropriate. The cause of this can be explained by the behaviour of the mean-reversion close to the unit root. Although the interval of the cross sectional mean reversion parameter, needed for extrapolation, is in between $0$ and $0.04$ this has large effects on $\theta$ and showed to be the parameter of concern determining the extrapolations. But the extreme uncertainty on $\theta$ does not add up in the extrapolations as for low mean-reversion the Vasicek model does not converge to $\theta$ within limited time spans while for larger mean-reversion the ultimate level is less uncertain. 

\vspace{5mm}
\noindent
According to the data the extrapolations contain very wide confidence intervals. If one believes that the uncertainty is much smaller, one indirectly claims to have more prior information at hand. Thus either we have to accept the problem of the size of the uncertainty or there is more information available that we are unaware of and which should be included in the priors to narrow the bounds. 

\vspace{5mm}
\noindent
Summarising, classical estimates lead to unreasonable (often negative) long term yields and extremely wide confidence intervals, but sensible Bayesian priors lead to more sensible extrapolations.

\newpage
\appendix
\newpage
\section*{Appendices}
\section{Relation $\mathbb{P}$ and $\mathbb{Q}$} 
\label{App:AppendixA} 
\setcounter{equation}{0}
Under the assumption that two short rate AR(1) models exist under two different probability measures, a risk-neutral $\mathbb{Q}$ and a risk-full $\mathbb{P}$ by the use of a stochastic discount factor (SDF) the relation between the parameters of the different measures can be derived as follows. 
Let the SDF be
\begin{equation}
\frac{d \Lambda}{\Lambda} = -r_t dt - \lambda dW_t \nonumber \end{equation} 
where 
\begin{equation} 
\lambda_t = \Lambda_0 + \Lambda_1 r_t \nonumber 
\end{equation} 
Since under the risk-neutral measure $\lambda=0$ we can derive the relation between the two probability measures. 
Here the continuous notation of the AR(1), the Vasicek model is used to come to the relationship.
\begin{eqnarray}
dr_t = \kappa (\mu - r_t) dt + \sigma dW_t 
\end{eqnarray} The log of the price will be affine with respect to the short rate $r$. 
Similar to the notation of Cochrane (2001)\nocite{cochrane2001asset}, where $T=t+\tau$ is the maturity date of the bond, and the price at $t$ is (thus $\tau$ is the remaining time to maturity) 
\begin{equation}
p(\tau,t,r) = -A(\tau)-B(\tau) r_t
\end{equation}
Thus the (antilog) of the price is (Duffie and Kan (1996))\nocite{duffie1996yield} 
\begin{equation}
P(\tau,t,r_t) = \textrm{exp}\left(-A(\tau)-B(\tau) r_t\right) 
\end{equation} 
By It$\bar{o}$'s Lemma 
\begin{eqnarray}
dP(\tau,t,r_t) &=& -B(\tau)Pdr_t +\left(\frac{\partial A(\tau)}{\partial t} +\frac{\partial B(\tau)}{\partial t}r_t \right)P dt + \frac{1}{2}B^2(\tau) \sigma^2 P dt \nonumber\\ \frac{dP(\tau,r_t)}{P} &=& -B(\tau)dr_t +\left(\frac{\partial A(\tau)}{\partial t} +\frac{\partial B(\tau)}{\partial t}r_t \right) dt + \frac{1}{2}B^2(\tau) \sigma^2 dt \nonumber 
\end{eqnarray} 
The Fundamental Pricing Equation states
\begin{eqnarray} 
&&\mathbb{E}_t\left[\frac{dP}{P}\right] - r_t dt = -\mathbb{E}\left[\frac{dP}{P}\frac{d\Lambda}{\Lambda}\right] \nonumber\\ 
&& -B(\tau)\kappa(\mu-r_t)dt +\left(\frac{\partial A(\tau)}{\partial t} +\frac{\partial B(\tau)}{\partial t}r_t \right) dt + \frac{1}{2}B^2(\tau) \sigma^2 dt - r_tdt = \nonumber\\ 
&&-B(\tau)\sigma \Lambda_0 dt - B(\tau) \sigma \Lambda_1 r_t dt \nonumber 
\end{eqnarray} 
Yields, ordered by all terms without $r$ 
\begin{eqnarray} 
&&-B(\tau)\kappa\mu dt  +\frac{\partial A(\tau)}{\partial t}dt + \frac{1}{2}B^2(\tau) \sigma^2 dt = -B(\tau)\sigma \Lambda_0 dt \nonumber \\
&& \Rightarrow \frac{\partial A(\tau)}{\partial t} = B(\tau) \left[\kappa \mu - \sigma \Lambda_0 \right]   - \frac{1}{2}B^2(\tau) \sigma^2 \nonumber
\end{eqnarray}
And the terms including $r$
\begin{eqnarray}
&& - B(\tau)\kappa r_t dt +\frac{\partial B(\tau)}{\partial t}r_t dt - rdt  = - B(\tau) \sigma \Lambda_1 r dt \nonumber \\ 
&& \Rightarrow \frac{\partial B(\tau)}{\partial t} = 1- B(\tau)\left[ \sigma \Lambda_1 + \kappa \right] \nonumber 
\end{eqnarray} 
For completeness the above formulas are all in terms of probability measure $\mathbb{P}$. The derivatives of component $A$ and $B$ are equal irrespective of the probability measure, hence we also know that under the risk-neutral measure $\lambda_t=0$,
\begin{eqnarray}
\frac{\partial A(\tau)}{\partial t} &=& B(\tau) \left[\kappa \mu - \sigma \Lambda_0 \right]   - \frac{1}{2}B^2(\tau) \sigma^2 \nonumber\\
&=& B(\tau) \left[\tilde{\kappa} \tilde{\mu} \right]   - \frac{1}{2}B^2(\tau) \sigma^2 \nonumber\\
\frac{\partial B(\tau)}{\partial t} &=& 1- B(\tau)\left[ \sigma \Lambda_1 + \kappa \right] \nonumber \\ 
&=& 1- B(\tau)\left[\tilde{\kappa} \right] \nonumber 
\end{eqnarray} 
\noindent 
If we put the terms in brackets equal to $\tilde{\kappa} \tilde{\mu}$ and $\tilde{\kappa}$ respectively we get 
\begin{eqnarray} 
\tilde{\kappa}= \kappa + \sigma \Lambda_1 \nonumber\\ 
\tilde{\mu} \tilde{\kappa} = \mu \kappa - \sigma \Lambda_0 
\end{eqnarray}

\newpage
\newpage
\section{Affine derivation of zero rates}
\label{App:AppendixB}
The Vasicek process of the short rate under $\mathbb{P}$ is 
\begin{equation}
dr_t = -\kappa(r_t-\mu)dt+ \sigma dW_t \nonumber
\end{equation}
By Ito's Lemma under $\mathbb{Q}$ this can be expressed directly
\begin{eqnarray}
r_{t+s} &=& r_t e^{-\tilde{\kappa} s} + \tilde{\mu} (1-e^{-\tilde{\kappa} s}) + \sigma \int_t^{t+s} e^{-\tilde{\kappa} (t+s-u)} dW_u \nonumber\\
\int_0^{\tau} r_{t+s} ds &=& \int_0^{\tau} r_t e^{-\tilde{\kappa} s}ds + \int_0^{\tau} \tilde{\mu} (1-e^{-\tilde{\kappa} s})ds + \int_0^{\tau} \int_t^{t+s} \sigma e^{-\tilde{\kappa} (t+s-u)} dW_u ds \nonumber\\  
&=& (r_t-\tilde{\mu}) \frac { (1-e^{-\tilde{\kappa} \tau} )}{\tilde{\kappa}} + \tilde{\mu} \tau  + \int_0^{\tau} \left( \int_{t}^{t+s} \sigma e^{-\tilde{\kappa} (t+s-u)} dW_u\right)ds \nonumber\\  
\end{eqnarray}
Change the order of integrals
\begin{eqnarray}
\int_0^{\tau} \left( \int_{t}^{t+s} \sigma e^{-\tilde{\kappa} (t+s-u)} dW_u\right)ds &=& \int_t^{t+\tau} \left( \int_{u-t}^{\tau} \sigma e^{-\tilde{\kappa} (t+s-u)} ds\right)dW_u  \nonumber\\
&=& \int_t^{t+\tau} \left( -\frac{\sigma}{\kappa}(1-e^{-\tilde{\kappa}(t+\tau-u)})\right)dW_u\nonumber
\end{eqnarray}
Let 
\begin{equation}
M = \mathbb{E}\bigg[\int_0^{\tau} r_{t+s} ds\bigg | r_t\bigg] = (r_t-\tilde{\mu}) \frac { (1-e^{-\tilde{\kappa} \tau} )}{\tilde{\kappa}} + \tilde{\mu} \tau
\end{equation}
and let
\begin{equation}
V = \textrm{var}\bigg[ \int_0^{\tau} r_{t+s} ds \bigg| r_t \bigg] = \frac{\sigma^2}{\tilde{\kappa}^2} \int_t^{t+\tau} (1- e^{-\tilde{\kappa} (t+\tau-u)})^2 du = \frac{\sigma^2}{\tilde{\kappa}^2} \left(\tau - \frac{1-e^{-\tilde{\kappa} \tau}}{\tilde{\kappa}} - \frac{{(1-e^{-\tilde{\kappa} \tau})}^2}{ 2 \tilde{\kappa}}\right)
\end{equation}
If $X \sim N(M, V)$ then $\mathbb{E}[e^X]= e^{M+\frac{1}{2}V}$
Thus
\begin{eqnarray}
\mathbb{E}[e^{-\int_0^{\tau} r_s ds}] &=& e^{-M+\frac{1}{2}V} = e^{-\tau z(\tau)} \nonumber\\
\tau z(\tau) &=& M-\frac{1}{2}V \nonumber\\
z(\tau) &=& (r_t-\tilde{\mu}) \frac { (1-e^{-\tilde{\kappa} \tau} )}{\tilde{\kappa} \tau } + \tilde{\mu} - \bigg( \frac{\sigma^2}{2\tilde{\kappa}^2} - \frac{\sigma^2}{2\tilde{\kappa}^2} \frac{1-e^{-\tilde{\kappa} \tau}}{\tilde{\kappa} \tau } - \frac{\sigma^2}{2\tilde{\kappa}^2} \frac{{(1-e^{-\tilde{\kappa} \tau})}^2}{ 2 \tilde{\kappa} \tau} \bigg) \nonumber
\end{eqnarray}
\noindent

\newpage
Let 
\begin{eqnarray}
{b}(\tau) &=& \frac{1-e^{-\tilde{\kappa} \tau}}{\tilde{\kappa} \tau} \nonumber\\
{\theta} &=& \tilde{\mu} -\frac{\sigma^2}{2\tilde{\kappa} ^2} \nonumber\\
\omega^ 2 &=& \frac{\sigma^2}{2\tilde{\kappa} } \nonumber
\end{eqnarray}
\noindent
Then it follows that
\begin{eqnarray}
z(\tau) &=& {b}(\tau) \bigg[r_t - {\theta} \bigg] + {\theta} + \frac{1}{2} \tau \omega^2 {{b}(\tau)}^2 \nonumber\\
z(s) &=& {b}(s) \bigg[r_t - {\theta} \bigg] + {\theta} + \frac{1}{2} s \omega^2 {{b}(s)}^2 \nonumber\\
z(s) &=& \frac{{b}(s)}{{b}(\tau)} \bigg[z(\tau) - {\theta} \bigg] + {\theta} + \frac{1}{2} \omega^2 {b}(s) \bigg(s {b}(s) - \tau {b}(\tau)\bigg) \nonumber
\end{eqnarray}

\newpage
\section{Affine relation short rates and zero rates}
\label{App:AppendixC}
The short rate follows a Vasicek model under the real world measure $\mathbb{P}$
\begin{equation}
dr_t = -\kappa(r_t-\mu)dt+ \sigma_r dW_t \nonumber
\end{equation}
The expectation implies a linear relation between the short rate $r$ and the zero rate $z$
\begin{equation}
\label{affine}
z(\tau) = {b}(\tau) \bigg[r_t - {\theta} \bigg] + {\theta} + \frac{1}{2} \tau {\omega}^2 {{b}(\tau)}^2
\end{equation}
Therefore the process $dz$ is also a Vasicek model. If we let
\begin{equation}
dz_t = -a(z_t-m)dt+ \sigma_z dW_t \nonumber
\end{equation}
we have to find the relations between the parameters of the two Vasicek models,
\begin{eqnarray}
\label{this}
dz(\tau) &=& d \left({b}(\tau)r_t - {b}(\tau) {\theta} + {\theta}+ \frac{1}{2} \tau {\omega}^2 {{b}(\tau)}^2 \right)\\
&=&  -{b}(\tau)\kappa(r-\mu)dt+ {b}(\tau)\sigma_r dW
\end{eqnarray}
Relation \eqref{affine}, here rewritten the other way round, for $r_t$ in terms of $z_t$
\begin{eqnarray}
\label{that}
r_t  &=& \frac{z(\tau)}{{b}(\tau)} -\frac{{\theta}}{{b}(\tau)} - \frac{1}{2} \tau {\omega}^2 {{b}(\tau)} + {\theta}
\end{eqnarray}
Plugging \eqref{that} into \eqref{this} leads to
\begin{equation}
dz_t(\tau)=  -\kappa\bigg(z(\tau) -{\theta} - \frac{1}{2} \tau {\omega}^2 {b}(\tau)^2+ {\theta} {b}(\tau)-\mu {b}(\tau) \bigg)dt+ {b}(\tau)\sigma_r dW \nonumber
\end{equation}
Thus
\begin{eqnarray}
a  &=& \kappa \nonumber\\
m &=& {\theta} + \frac{1}{2} \tau {\omega}^2 {b}(\tau)^2 - {\theta}{b}(\tau) + \mu {b}(\tau) \nonumber\\
\sigma_z &=& {b}(\tau)\sigma_r \nonumber
\end{eqnarray}

\newpage
\section{Covariance decomposition}
\label{App:AppendixD}

\begin{table}[here]
\caption{Relations per decomposition}
\begin{center}
  \begin{tabular}{ l l l  }
    \hline
     & Corr  & Noise \\
     \hline
    $\tilde{\kappa}$ &  $\frac{\displaystyle 1-\textrm{e}^{-\tilde{\kappa}\tau_1}}{\displaystyle 1-\textrm{e}^{-\tilde{\kappa}\tau_2}} = \sqrt{\frac{\displaystyle \sigma_{(11)}}{\displaystyle \sigma_{(22)}}} \frac{\displaystyle \tau_1}{\displaystyle \tau_2} $ &  $\frac{\displaystyle (1-\textrm{e}^{-\tilde{\kappa}\tau_1})\tau_2}{\displaystyle (1-\textrm{e}^{-\tilde{\kappa}\tau_2})\tau_1} - \frac{\displaystyle (1-\textrm{e}^{-\tilde{\kappa}\tau_2})\tau_1}{\displaystyle (1-\textrm{e}^{-\tilde{\kappa}\tau_1})\tau_2} = \frac{\displaystyle \sigma_{(11)}-\sigma_{(22)}}{\displaystyle \sigma_{(21)}}$ \\
            $\sigma^2$ & $\frac{\displaystyle \sigma_{(11)}}{\displaystyle \tilde{b}(\tau_1)^2}$, $ \frac{\displaystyle \sigma_{(22)}}{\displaystyle \tilde{b}(\tau_2)^2}$ & $\frac{\displaystyle \sigma_{(21)}}{\displaystyle {b}(\tau_1){b}(\tau_2)} $ \\
        $\rho$ &  $\frac{\displaystyle \sigma_{(21)}}{\displaystyle \sqrt{\sigma_{(11)}\sigma_{(22)}}}$ & -  \\
      $\eta$ & - & ${\sigma_{(11)}-\sigma^2 {b}(\tau_1)^2}$, ${\sigma_{(22)}-\sigma^2 {b}(\tau_2)^2}$ \\
    \hline
  \end{tabular}
\end{center}
\label{table:corr_noise}
\end{table}
\noindent

\begin{table}[here]
\caption{Common relations}
\begin{center}
\begin{tabular}{l l}
\hline
& Corr and Noise \\
\hline
            $\kappa$ & $- \frac{\displaystyle \ln(1-a^dh)}{\displaystyle h}$ \\  
             $\mu$ & $\frac{\displaystyle (1-{b}(\tau_2))m(\tau_1)-(1-{b}(\tau_1))m(\tau_2)}{\displaystyle {b}(\tau_1)-{b}(\tau_2)}- \frac{\displaystyle 1}{\displaystyle 2} {{\omega}^2} \frac{\displaystyle \tau_1 {b}(\tau_1)^2(1-{b}(\tau_2)) - \tau_2 {b}(\tau_2)^2(1-{b}(\tau_1))}{\displaystyle {b}(\tau_1)-{b}(\tau_2)}$ \\
            $\tilde{\mu}$ & ${{\theta}} + \frac{\displaystyle \sigma^2}{\displaystyle 2\tilde{\kappa}^2} $ \\
        ${\theta}$ & $\frac{\displaystyle {b}(\tau_2)m(\tau_1)-{b}(\tau_1)m(\tau_2)}{\displaystyle {b}(\tau_2)-{b}(\tau_1)} - \frac{\displaystyle 1}{\displaystyle 2} {{\omega}^2 {b}(\tau_1){b}(\tau_2)} \frac{\displaystyle \tau_1 {b}(\tau_1) - \tau_2 {b}(\tau_2)}{\displaystyle {b}(\tau_2)-{b}(\tau_1)}$ \\
     $\Lambda_0$ & $\frac{\displaystyle \mu \kappa - \tilde{\mu} \tilde{\kappa}}{\displaystyle \sigma}$ \\
        $\Lambda_1$ & $\frac{\displaystyle \tilde{\kappa}-\kappa}{\displaystyle \sigma}$ \\
    \hline
\end{tabular}
\end{center}
\label{table:common_corr_noise}
\end{table}
\noindent

\newpage
\section{Conditional Maximum Likelihood Estimators}
\label{App:AppendixE}
If we do not consider a Bayesian approach but use a frequentist approach we start with the same likelihood function based on
\begin{equation}
\boldsymbol{Z}_t=\boldsymbol{Z}_{t-h}- a h(\boldsymbol{Z}_{t-h} - \boldsymbol{m})+\sqrt{h} \boldsymbol{\sigma} e_t \nonumber
\end{equation}
where $e_t^{(1)}$ and $e_t^{(2)}$ are from a bivariate standard Normal distribution.

\vspace{5mm}
\noindent
The likelihood function is
\begin{eqnarray}
L(\boldsymbol{Z}|\boldsymbol{m}, a,\boldsymbol{\varSigma}) &=& \left(2 \pi h|\boldsymbol{\varSigma}|\right)^{-N/2} \textrm{exp}\bigg(-\frac{1}{2h} (\boldsymbol{Z}_t-\boldsymbol{Z}_{t-h}+ a h{\boldsymbol{Z}_{t-h}}- ah { \boldsymbol{\iota m}}')'\nonumber\\
&&(\boldsymbol{Z}_t-\boldsymbol{Z}_{t-h}+ a h{\boldsymbol{Z}_{t-h}}- ah { \boldsymbol{\iota m}}')  \boldsymbol{\varSigma}^{-1}\bigg) \nonumber
\end{eqnarray}
where the dimensions are, $ \boldsymbol{Z}_t  = [N \times 2], \boldsymbol{m}  = [2 \times 1], a = [1 \times 1], \boldsymbol{\varSigma}  = [2 \times 2]$
$ h = [1 \times 1], \boldsymbol{\iota} = [N \times 1]$.

\vspace{5mm}
\noindent
We can apply the conditional Maximum Likelihood Estimation (cMLE) and get estimators for $\kappa$ and $\mu$, instead of the MCMC method by using Bayes' theorem. Note that we condition on the first observation, the cMLE works under the assumption that $z_0$ is given. As a mathematical convention the marginal distribution of the initial starting point is assumed to be a Dirac Delta function approaching one, which can be seen as the limit of a Normal where the uncertainty disappears. Under a classical interpretation asymptotically the parameters are Normal distributed with a mean equal to the cMLE and the variance obtained by the inverse of the negative expectation of the second order derivative. The difference with the Bayesian approach is that this distribution is achieved based on the assumption of repeated sampling, whereas a Bayesian approach is conditioned on a finite sample which makes is  valuable in a limited number of data-points as is often the case in term-structure models.

\vspace{5mm}
\noindent
The conditional log-likelihood is
\begin{eqnarray}
\ell= \log L(\boldsymbol{Z}|\boldsymbol{m}, a,\boldsymbol{\varSigma}) &\propto&   - \frac{N}{2}\log{|\boldsymbol{\varSigma}|} -\frac{1}{2 h} (\boldsymbol{Z}_t-\boldsymbol{Z}_{t-h}+ a h\boldsymbol{Z}_{t-h}- ah \boldsymbol{\iota m}')'\nonumber\\
&&(\boldsymbol{Z}_t-\boldsymbol{Z}_{t-h}+ a h\boldsymbol{Z}_{t-h}- ah \boldsymbol{\iota m}')  \boldsymbol{\varSigma}^{-1} 
\end{eqnarray}

\noindent
The solutions of the maximum likelihood of the parameters $a, \boldsymbol{m}$ and $\boldsymbol{\varSigma}$ are the same as the ones that minimize the ordinary least squares. Hence we get
\begin{eqnarray}
a_{\textrm{cMLE}} &=& \frac{\textrm{tr}( (\boldsymbol{\iota m}' - \boldsymbol{Z}_{t-h})'\boldsymbol{\Delta Z})}{\textrm{tr}( h (\boldsymbol{\iota m}' - \boldsymbol{Z}_{t-h})'(\boldsymbol{\iota m}' - \boldsymbol{Z}_{t-h}))} \nonumber\\
\boldsymbol{m}_{\textrm{cMLE}} &=& \frac{\left( \boldsymbol{\Delta Z} + a h { \boldsymbol{Z}_{t-h}} \right)'\boldsymbol{\iota}}{a h N } \nonumber\\
\boldsymbol{\varSigma}_{\textrm{cMLE}} &=&  \frac{(\boldsymbol{Z}_t-\boldsymbol{Z}_{t-h}+ a h{\boldsymbol{Z}_{t-h}}- ah \boldsymbol{\iota m}')'(\boldsymbol{Z}_t-\boldsymbol{Z}_{t-h}+ a h\boldsymbol{Z}_{t-h}- ah \boldsymbol{\iota m}')}{h N}\nonumber
\end{eqnarray}

\vspace{5mm}
\noindent
Now if we take the second order derivates of the conditional log-likelihood we can derive the asymptotic variances. The second order derivative results in a matrix, which is the information matrix if we take the negative expectation. The inverse yields the asymptotic covariance matrix for $a, \boldsymbol{m}$. 
\noindent
Thus the asymptotic distributions are
\begin{eqnarray}
a &\sim\!\!\!\!\!\!\!^{{}^{\textrm{asy}}}& N\left(a_{\textrm{cMLE}},\left\{\textrm{tr}\left(\boldsymbol{\varSigma}^{-1}_{\textrm{cMLE}}h \left( \boldsymbol{\iota m}_{\textrm{cMLE}}' -\boldsymbol{Z}_{t-h}\right)' \left( \boldsymbol{\iota m}_{\textrm{cMLE}}'-  \boldsymbol{Z}_{t-h}\right)\right)\right\}^{-1}\right) \nonumber\\
\boldsymbol{m} &\sim\!\!\!\!\!\!\!^{{}^{\textrm{asy}}}& N_2\left(\boldsymbol{m}_{\textrm{cMLE}},\left\{\boldsymbol{\varSigma}_{\textrm{cMLE}}^{-1} N a_{\textrm{cMLE}}^2 h\right \}^{-1}\right) \nonumber
\end{eqnarray}

\noindent
The asymptotic variance of $\boldsymbol{\varSigma}_{\textrm{cMLE}}$ is obtained by the usual procedure on the Hessian matrix (the inverse of minus the expecatation of the diagonal entries) of the $\textrm{vech}(\boldsymbol{\varSigma}_{\textrm{cMLE}})$ (Magnus and Neudecker (1988)\nocite{magnus1988matrix}). 
\begin{eqnarray}
\textrm{vech}(\boldsymbol{\varSigma}_{\textrm{cMLE}}) = (\sigma_{(11)},\sigma_{(21)},\sigma_{(22)})'\nonumber
\end{eqnarray}
Since $\sigma_{(12)}=\sigma_{(21)}$, there exists a premultiplication by the duplication matrix $\boldsymbol{D}_2$ equating to $\textrm{vec}(\boldsymbol{\varSigma}_{\textrm{cMLE}})$. Let $\boldsymbol{D}^{+}_2 = (\boldsymbol{D}_2'\boldsymbol{D}_2)^{-1} \boldsymbol{D}_2'$.
Then Magnus and Neudecker give the general result of the variance of the half vectorization, resulting in the asymptotic distribution 
\begin{eqnarray}
\textrm{vech}\left(\boldsymbol{\varSigma}\right) &\sim\!\!\!\!\!\!\!^{{}^{\textrm{asy}}}& N_2\left(\textrm{vech}\left(\boldsymbol{\varSigma}_{\textrm{cMLE}}\right),2 \boldsymbol{D}^{+}_2 \left(\boldsymbol{\varSigma}_{\textrm{cMLE}} \otimes \boldsymbol{\varSigma}_{\textrm{cMLE}}\right) \boldsymbol{D}^{+}_2\right) \nonumber
\end{eqnarray}

\newpage
\section{Hyperparameters conditional posterios}
\label{App:AppendixC}
 
 The conditional posterior distributions are respectively,
 \begin{eqnarray}
 f(a |\boldsymbol{Z},\boldsymbol{m},\boldsymbol{\varSigma}) &\propto& f(\boldsymbol{Z}|\boldsymbol{z}_0,a,\boldsymbol{m},\boldsymbol{\varSigma}) f(a) \nonumber\\ 
 &\sim& N\bigg( \mu_{ca}(\boldsymbol{Z},\boldsymbol{m},\boldsymbol{\varSigma}) , \tau_{ca}(\boldsymbol{Z},\boldsymbol{m},\boldsymbol{\varSigma)}\bigg) \nonumber\\ 
 f(\boldsymbol{m} |\boldsymbol{Z},a,\boldsymbol{\varSigma}) &\propto& f(\boldsymbol{Z}|\boldsymbol{z}_0,a,\boldsymbol{m},\boldsymbol{\varSigma}) f(\boldsymbol{m}) \nonumber\\ 
 &\sim& N_2\bigg( \boldsymbol{\mu_{cm}}(a,\boldsymbol{\varSigma},\boldsymbol{Z}) , \boldsymbol{\Omega_{cm}}(a,\boldsymbol{\varSigma},\boldsymbol{Z})\bigg)\nonumber\\ 
 f(\boldsymbol{\varSigma}^{-1}|\boldsymbol{Z},a,\boldsymbol{m}) &\propto& f(\boldsymbol{Z}|\boldsymbol{z}_0,a,\boldsymbol{m},\boldsymbol{\varSigma})f(\boldsymbol{\varSigma}^{-1}) \nonumber\\
 &\sim& W_2\bigg(\boldsymbol{\Psi_{c\varSigma}}(\boldsymbol{Z},a,\boldsymbol{m}), \nu_{c\boldsymbol{\varSigma}}(\boldsymbol{Z},a,\boldsymbol{m})\bigg) \nonumber
 \end{eqnarray}
 where subscript $a,\boldsymbol{m}$ or $\boldsymbol{\varSigma}$ denote the prior means and (co)variances and the $c$'s in front denote the conditional posterior means and (co)variances. These posterior hyperparameters are all functions dependent on the other parameters that is being conditioned on.
 \vspace{5mm}
 \noindent
 \begin{eqnarray}
 \mu_{ca} &=&  \left(\textrm{tr}(\boldsymbol{\varSigma}^{-1} \boldsymbol{\Delta Z}'(\boldsymbol{\iota} \boldsymbol{m}' - \boldsymbol{Z}_{t-h})) + \frac{\mu_{a}}{\tau_{a}^2}\right) \nonumber\\
 && \cdot \left(h \textrm{tr}(\boldsymbol{\varSigma}^{-1}(\boldsymbol{\iota} \boldsymbol{m}' - \boldsymbol{Z}_{t-h})'(\boldsymbol{\iota} \boldsymbol{m}' - \boldsymbol{Z}_{t-h})) + \frac{1}{\tau_{a}^{2}}\right) \nonumber\\
 \tau_{ca}^{-2} &=& h \textrm{tr}(\boldsymbol{\varSigma}^{-1}(\boldsymbol{\iota} \boldsymbol{m}' - \boldsymbol{Z}_{t-h})'(\boldsymbol{\iota} \boldsymbol{m}' - \boldsymbol{Z}_{t-h})) + \frac{1}{\tau_{a}^{2}} \nonumber\\
 \boldsymbol{\mu_{cm}} &=& \left(\boldsymbol{\Omega_m}^{-1} + a^2  T \boldsymbol{\varSigma}^{-1} \right)^{-1} \cdot \left(\boldsymbol{\Omega_m}^{-1} \boldsymbol{\mu_m} + a \boldsymbol{\varSigma}^{-1} \boldsymbol{\iota}'(\boldsymbol{\Delta Z}_t+ a h{\boldsymbol{Z}_{t-h}}) \right) \nonumber\\
 \boldsymbol{\Omega_{cm}}^{-1} &=& 
 \left(\boldsymbol{\Omega_m}^{-1} + a^2  T\boldsymbol{\varSigma}^{-1} \right) \nonumber\\
 \boldsymbol{\Psi_{c\varSigma}} &=& \boldsymbol{\Psi_{\varSigma}}+ h^{-1 }(\boldsymbol{\Delta Z_t}+ a h{\boldsymbol{Z}_{t-h}}-ah {\boldsymbol{\iota} \boldsymbol{m}'})'(\boldsymbol{\Delta Z}_t+ a h{\boldsymbol{Z}_{t-h}}-ah {\boldsymbol{\iota m}'}) \nonumber\\
 \nu_{c\boldsymbol{\varSigma}} &=& \nu_{\boldsymbol{\varSigma}}  + N \nonumber
 \end{eqnarray}

\newpage
\section{Maximum Likelihood extrapolation}

\noindent
\begin{figure}[!h]
\centering
\caption*{cMLE extrapolation }
\begin{subfigure}[b]{\textwidth} 
	\centering 
	\includegraphics[width=\textwidth]{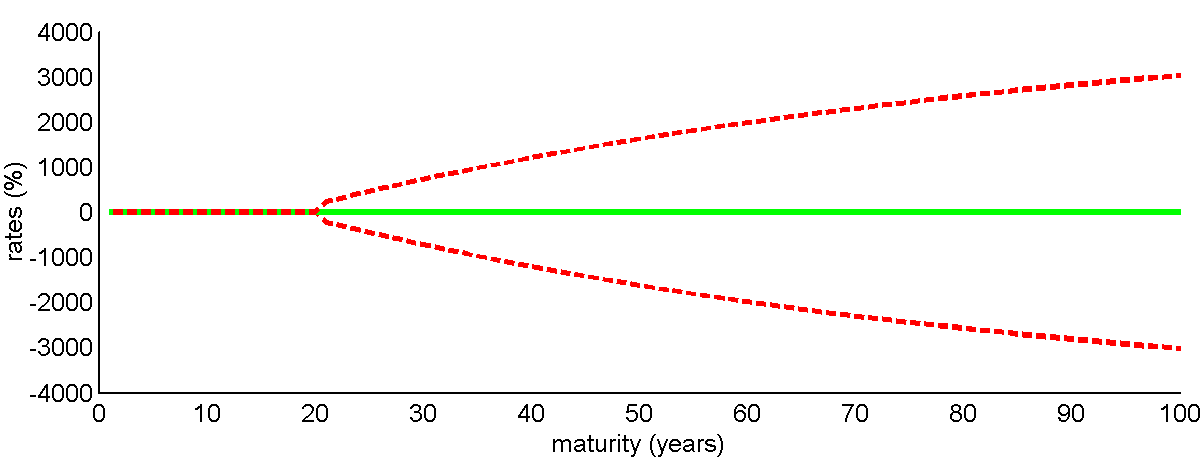}
	\caption{All simulations} 
\end{subfigure}
~
\begin{subfigure}[b]{\textwidth}
	\centering
	\includegraphics[width=\textwidth]{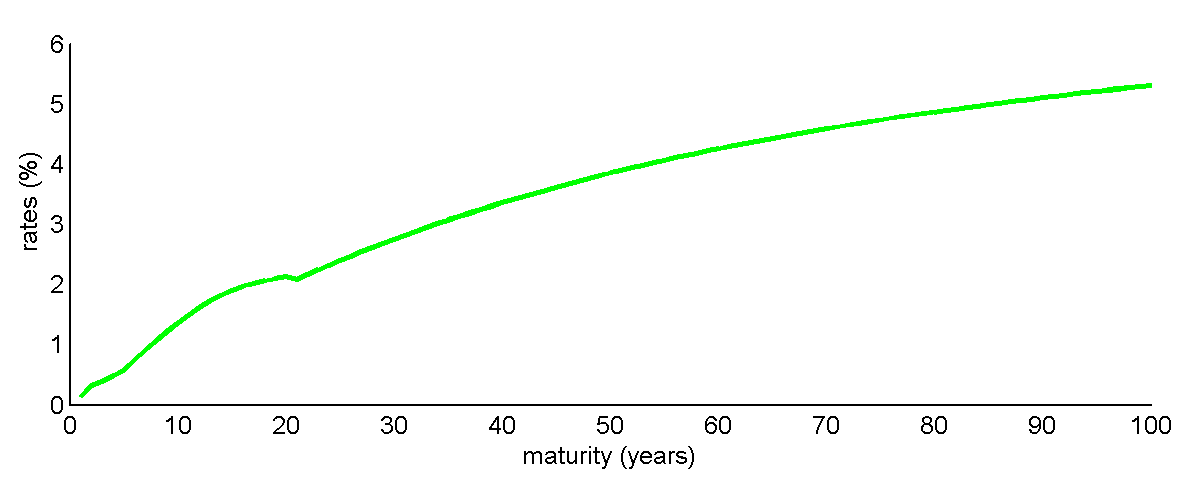}
	\caption{Average}
\end{subfigure}
\caption{\small{\textit{Based on last liquid point of 20-year maturity from September 2013, $\tau_1=5$ and $\tau_2=20$. The dashed red line is the $95\%$ confidence interval, and the green line the point estimate for the maturity ranging from $21$ to $100$ years.}}}
\end{figure}

\newpage
\section{Robustness check}
\label{App:AppendixG}
Similar results for different choice of cross-sectional data.
Noise for $\tau_1$ =       10, $\tau_2$ =       20.

\begin{table}[here] 
\caption{Corr for $\tau_1$ =       10, $\tau_2$ =       20} 
\begin{center} 
  \begin{tabular}{ l l l l l l l} 
    \hline 
& Average & HPD95 lb & HPD95 ub & CI95 lb & CI95 ub & St. Dev.\\ 
\hline 
$\kappa$ &   0.1358 &    4.433e-006 &   0.2867 &    1.223e-002 &   0.3291 &   0.0822 \\ 
 $\tilde{\kappa}$ &   0.0057 &    1.482e-008 &   0.0135 &    2.486e-004 &   0.0154 &   0.0041\\ 
 $\mu$ &   0.0068  &   -1.130e-002 &   0.0301 &   -1.018e-002 &   0.0316  &   0.0149\\ 
 $\tilde{\mu}$ & 143.2251   &   -2.152e-001 &   4.5827 &    5.887e-002 &   8.6096 & 140595.8522\\ 
 $ {\theta}$ &   -2.195e+005    & -87.6128 &   0.7061 &  -346.9505 &   0.1584 &   1.070e+008 \\ 
 $\Lambda_0$ &   -0.1474 & -0.9039   &   0.7511 & -0.8099  &   0.8824 &   0.5561\\ 
 $\Lambda_1$ & -18.8848 & -41.3921  &   0.8432 & -47.0027 &  -0.9218 &  11.9808 \\ 
 $\sigma^2$ &    4.808e-005 &    3.642e-005  &    6.094e-005 &    3.733e-005 &    6.230e-005 &    6.386e-006 \\ 
 $\rho$ &   0.9350 &   0.9137 &   0.9549 &   0.9118 &   0.9535 &   0.0106 \\ 
    \hline 
\end{tabular} 
\end{center} 
\end{table} 
\noindent 

\begin{table}[here] 
\caption{Noise for $\tau_1$ =       10, $\tau_2$ =       20} 
\begin{center} 
  \begin{tabular}{ l l l l l l l} 
    \hline 
& Average & HPD95 lb & HPD95 ub &CI95 lb & CI95 ub & St. Dev.\\ 
\hline 
$\kappa$ &   0.1358 &    4.433e-006 &   0.2867 &    1.223e-002 &   0.3291 &   0.0822 \\ 
 $\tilde{\kappa}$ &   0.0061 &    1.591e-008 &   0.0145 &    2.657e-004 &   0.0166 &   0.0044\\ 
 $\mu$ &   0.0069 &     -1.130e-002 &   0.0302 &   -1.014e-002 &   0.0317  &   0.0114\\ 
 $\tilde{\mu}$ &  59.0114   &   -2.053e-001 &   4.2427 &    5.219e-002 &   7.9482 & 47266.9622\\ 
 ${\theta}$ &   -1.796e+005 &    -71.3765 &   0.7569 &  -283.6703 &   0.1875 &   8.694e+007 \\ 
 $\Lambda_0$ &   -0.1476 &  -0.9337  & 0.7778   &  -0.8333 &   0.9163 &   0.4414\\ 
 $\Lambda_1$ & -19.4179 & -42.6660  &   0.9565 & -48.4647 &  -0.8835 &  12.3684 \\ 
 $\sigma^2$ &    4.527e-005 &    3.350e-005  &    5.802e-005 &    3.452e-005 &    5.950e-005 &    6.388e-006 \\ 
 $\eta$ &    2.828e-006 &    2.205e-006 &    3.515e-006 &    2.241e-006 &    3.568e-006 &    3.393e-007 \\ 
    \hline 
\end{tabular} 
\end{center} 
\end{table} 
\noindent

\newpage
\noindent
Different hyperparameters for the prior of $\boldsymbol{m}$
\begin{eqnarray}
\boldsymbol{\mu_m}&=&\begin{bmatrix} 
-1.1 \\
-1.1
\end{bmatrix} \nonumber\\
\boldsymbol{\Omega_m} &=& 
\begin{bmatrix}
0.3^2 & 0 \\
0 & 0.3^2
\end{bmatrix} \nonumber
\end{eqnarray}

\noindent
The truncated expectation and variance are 
\begin{eqnarray}
\mathbb{E}[m(\tau_1)]  &=& 0.0727 \nonumber\\
\textrm{var}[m(\tau_1)]  &=& 0.0692 \nonumber
\end{eqnarray}

\noindent
Noise for $\tau_1$ =        5, $\tau_2$ =       20
\begin{table}[here] 
\caption{Corr for $\tau_1$ =        5, $\tau_2$ =       20} 
\begin{center} 
  \begin{tabular}{ l l l l l l l} 
    \hline 
& Average & HPD95 lb & HPD95 ub & CI95 lb & CI95 ub & St. Dev.\\ 
\hline 
$\kappa$ &   0.1632 &    1.675e-005 &   0.3281 &    1.607e-002 &   0.3641 &   0.0922 \\ 
 $\tilde{\kappa}$ &   0.0156 &    1.317e-003 &   0.0289 &    2.538e-003 &   0.0305 &   0.0072\\ 
 $\mu$ &   0.0111  &   -7.588e-003 &   0.0289 &   -5.858e-003 &   0.0317  &   0.0094\\ 
 $\tilde{\mu}$ &   0.3143   &   -1.049e-002 &   0.6759 &    5.098e-002 &   1.0089 &   1.8332\\ 
 $\tilde{\theta}$ &   -1.157e+001    &  -0.9937 &   0.3674 &   -2.9600 &   0.1959 &   6.652e+002 \\ 
 $\Lambda_0$ &   -0.1251 &  -0.7292  &   0.5241 &  -0.6731 &   0.6014 &   0.3057\\ 
 $\Lambda_1$ & -19.3512 & -41.8920  &   1.8826 & -46.0028 &   0.0252 &  12.1868 \\ 
 $\sigma^2$ &    5.916e-005 &    4.408e-005  &    7.573e-005 &    4.522e-005 &    7.740e-005 &    8.244e-006 \\ 
 $\rho$ &   0.7731 &   0.7052 &   0.8368 &   0.7006 &   0.8332 &   0.0339 \\ 
    \hline 
\end{tabular} 
\end{center} 
\end{table} 
\noindent 

\begin{table}[here] 
\caption{Noise for $\tau_1$ =        5, $\tau_2$ =       20} 
\begin{center} 
  \begin{tabular}{ l l l l l l l} 
    \hline 
& Average & HPD95 lb & HPD95 ub &CI95 lb & CI95 ub & St. Dev.\\ 
\hline 
$\kappa$ &   0.1632 &    1.675e-005 &   0.3281 &    1.607e-002 &   0.3641 &   0.0922 \\ 
 $\tilde{\kappa}$ &   0.0204 &    1.683e-003 &   0.0383 &    3.303e-003 &   0.0407 &   0.0096\\ 
 $\mu$ &   0.0111 &     -7.646e-003 &   0.0290 &   -5.923e-003 &   0.0318  &   0.0094\\ 
 $\tilde{\mu}$ &   0.2399   &   -7.109e-003 &   0.5145 &    3.957e-002 &   0.7578 &   1.3671\\ 
 $\tilde{\theta}$ &   -5.278e+000 &     -0.3579 &   0.3212 &   -1.1267 &   0.2095 &   3.102e+002 \\ 
 $\Lambda_0$ &   -0.1444 &  -0.8182  &   0.5871 &  -0.7628 &   0.6633 &   0.3415\\ 
 $\Lambda_1$ & -20.7545 & -46.0823  &   2.8917 & -50.6134 &   0.7909 &  13.6014 \\ 
 $\sigma^2$ &    4.860e-005 &    3.326e-005  &    6.558e-005 &    3.435e-005 &    6.722e-005 &    8.419e-006 \\ 
 $\eta$ &    1.085e-005 &    8.427e-006 &    1.345e-005 &    8.602e-006 &    1.369e-005 &    1.301e-006 \\ 
    \hline 
\end{tabular} 
\end{center} 
\end{table} 
\noindent 

\section{Convergence tests}
\label{App:AppendixI}

\vspace{5mm}
\noindent
The standardized CUSUM statisic (Page (1954))\nocite{page1954continuous} for scalar $\theta$ is
\begin{equation}
CS_t = \frac{\frac{1}{t} \sum_{i=1}^{t} \theta^{(i)} - m_\theta}{s_\theta}
\end{equation}
where $m_\theta$ and $s_\theta$ are the MC sample mean and standard deviation of the $n$ draws.
If the MCMC sampler converges, the graph of the $CS_t$ against $t$ should converge smoothly to zero. On the contrary, long and regular excursions away from zero are an indication of the absence of convergence. 

\vspace{5mm}
\noindent
Geweke's test (Geweke et al. (1991))\nocite{geweke1991evaluating} 
compares the estimate of $\overline{g}_A$ of a posterior mean from the first $n_A$ draws with the estimate $\overline{g}_B$ from the last $n_B$ draws. If the two subsamples are well separated (i.e. there are many observations between them), they should be independent. The statistic is
\begin{equation}
Z = \frac{\overline{g}_A - \overline{g}_B}{(nse_A^2 + nse_B^2)^{1/2}}
\end{equation}
where $nse_A$ and $nse_B$ are the numerical standard errors of each subsample,
is normally distributed if $n$ is large and the chain has converged.
For a critical value of $5\%$ we do not reject the null, where the null states that the two subsamples deviate too much from each other. 

\vspace{5mm}
\noindent
Also the graphs of the autocorrelations give an indication whether the data the data is independent.

\begin{table}[here] 
\caption{Convergence tests}\begin{center} 
  \begin{tabular}{ l l l l l l l} 
    \hline 
& $a$ & $m(\tau_1)$ & $m(\tau_2)$ & $\sigma_{(11)}$ & $\sigma_{(21)}$ & $\sigma_{(22)}$ \\ 
\hline 
Geweke test &        0 &        0 &        0 &        0 &        0 &        0 \\ 
 Geweke Z &  -0.2740 &   1.5190 &   1.6329 &  -1.0051 &  -0.9190 &  -1.1027\\ 
 CUSUM mean  &     1.633e-01 &       1.454e-02 &     2.802e-02 &     5.421e-05 &     3.750e-05  &     4.329e-05\\ 
 CUSUM variance &     7.198e-03   &     5.019e-05 &     5.559e-05 &     4.203e-11 &     2.728e-11 &     2.657e-11\\ 
    \hline 
\end{tabular} 
\end{center} 
\end{table} 
\noindent

\noindent
\begin{figure}[h!]
\centering
\caption*{ACF test}
\begin{subfigure}[b]{0.73\textwidth}
	\centering
	\includegraphics[width=\textwidth]{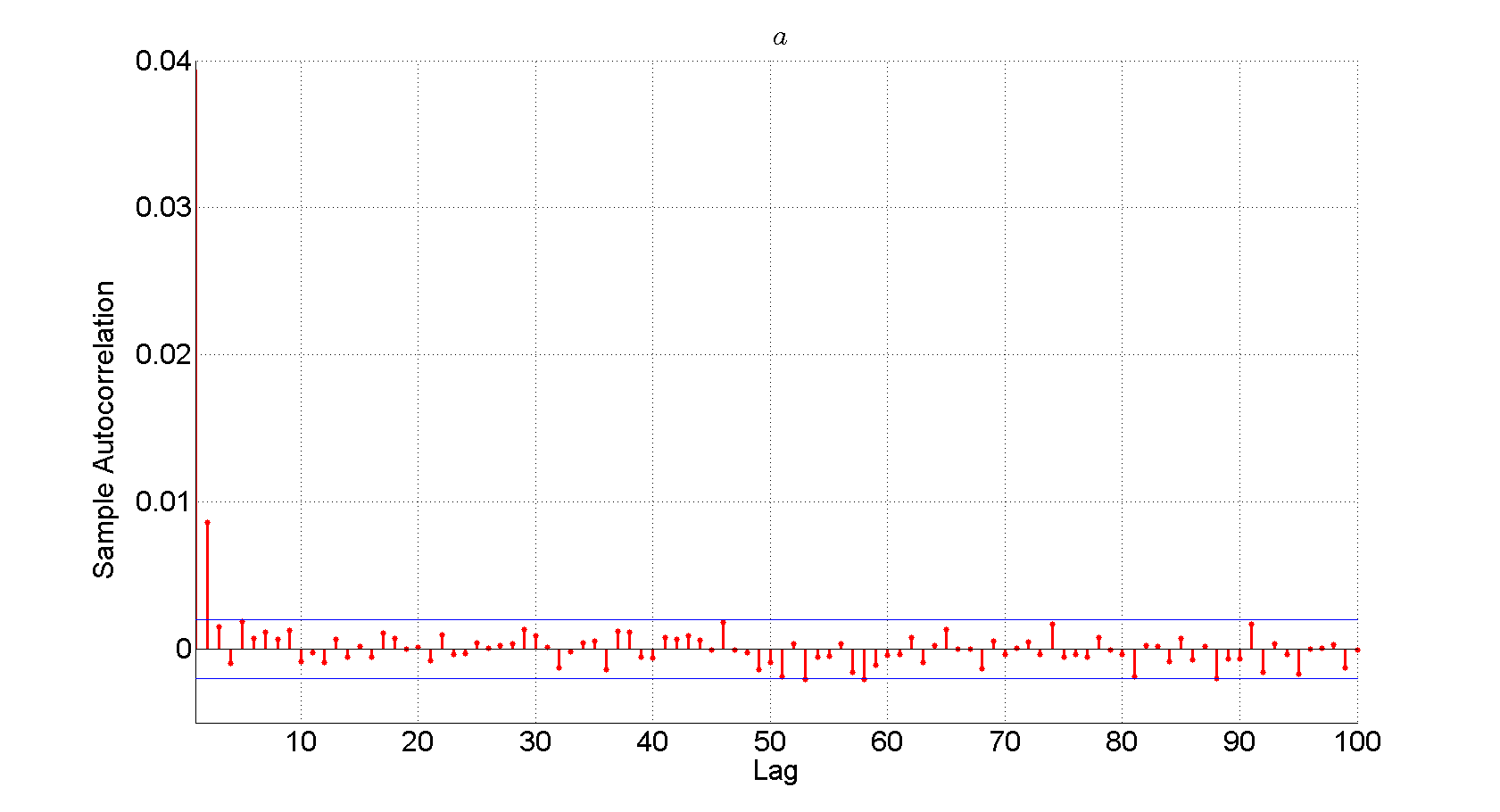}
\end{subfigure}
\caption{{\small \textit{ACF tests for the draws of $a$ for lags $1,2,...,100$. Similar results for $\boldsymbol{m}, \boldsymbol{\varSigma}$. All indicating a quick decline in the autocorrelations dependence. By construction the dependence of the zero lag is 1 and left out for visual convenience.}}}\label{fig:acf}
\end{figure}

\noindent
\begin{figure}[h!]
\centering
\caption*{CUSUM test}
\begin{subfigure}[b]{0.73\textwidth}
	\centering
	\includegraphics[width=\textwidth]{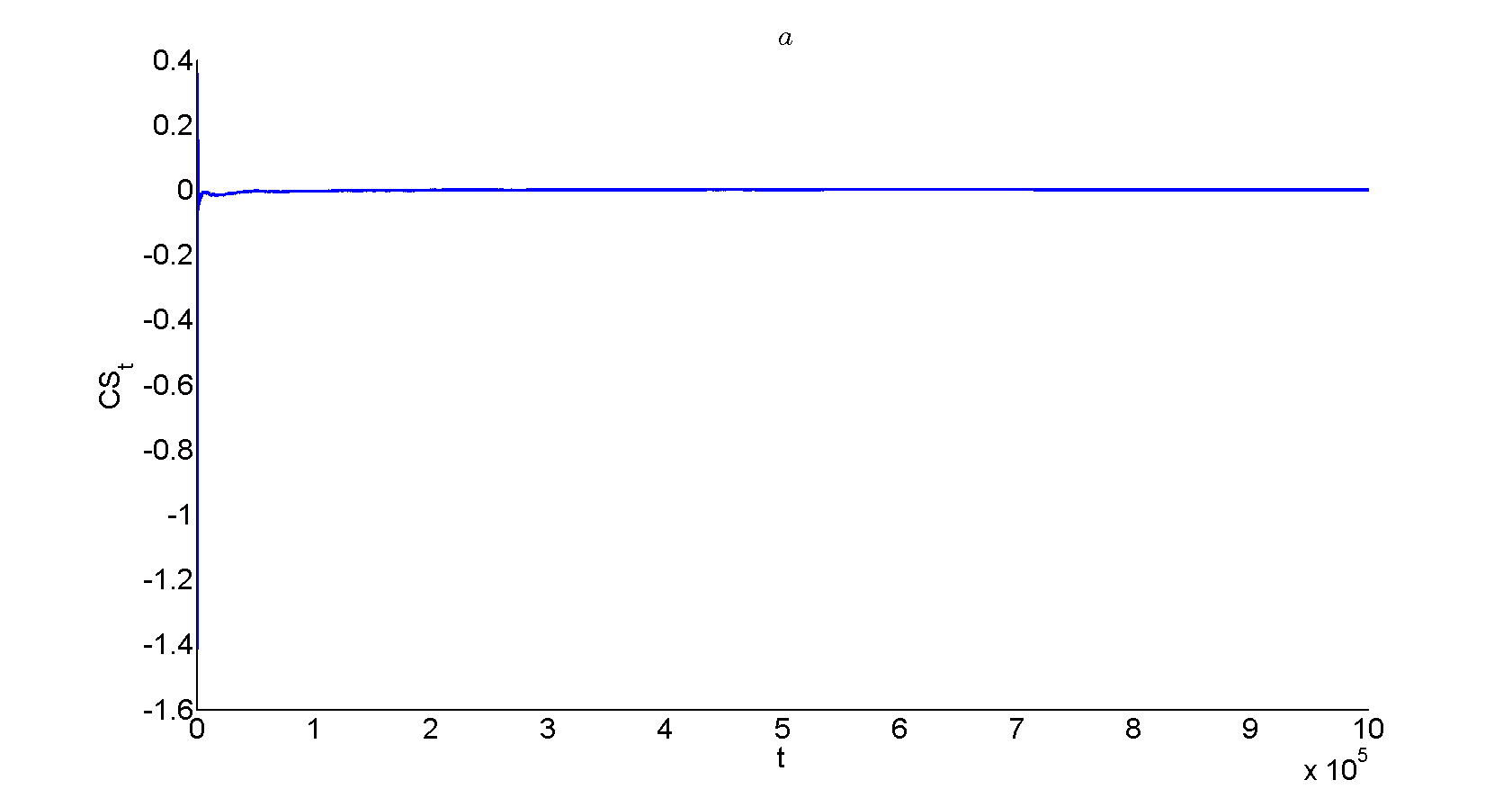}
\end{subfigure}
~
\begin{subfigure}[b]{0.73\textwidth}
	\centering
	\includegraphics[width=\textwidth]{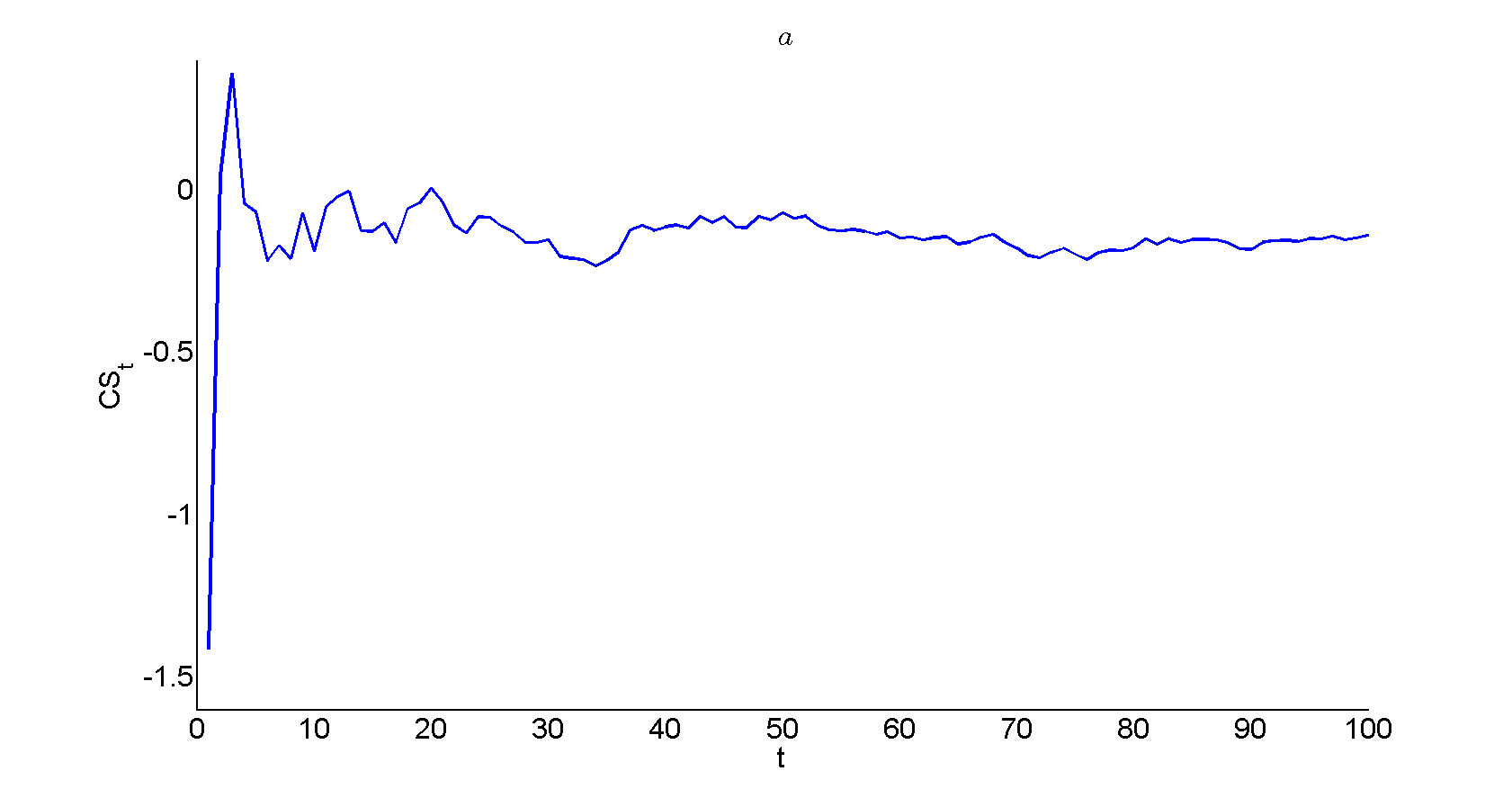}
\end{subfigure}
\caption{{\small \textit{The CUSUM test for the all the 1,000,000 draws of $a$ and a zoom of the first 100 draws are shown. Similar results hold for $\boldsymbol{m}, \boldsymbol{\varSigma}$. All show a quick convergence to zero, indicating that the overall average is achieved within the sample size.}}}\label{fig:cumsum}
\end{figure}

\newpage
\clearpage
\bibliographystyle{plain}

\bibliography{bibliography_trial}

\end{document}